\documentclass[ twocolumn, astrosymb]{aastex631}

\newcommand{\HI}{H\,\textsc{i}}
\newcommand{\HII}{H\,\textsc{ii}}
\newcommand{\NII}{N\,\textsc{ii}}
\newcommand{\SII}{S\,\textsc{ii}}
\newcommand{\OIII}{O\,\textsc{iii}}
\newcommand{\Ha}{H{$\alpha$}}
\newcommand{\Hb}{H{$\beta$}}

\received{August 29, 2025}
\revised{January 23, 2026}
\accepted{April 17, 2026}

\submitjournal{ApJ}

\begin{document}

\title{MAUVE--MUSE: Ionization and Kinematic Signatures of Environmental Effects on Virgo Cluster Disks}

\shorttitle{MAUVE--MUSE: Ionization and Kinematic Signatures in Virgo}

\correspondingauthor{Toby Brown}
\email{tobias.brown@nrc-cnrc.gc.ca}

\author[0000-0003-1845-0934]{Toby Brown}
\affiliation{National Research Council of Canada, 
Herzberg Astronomy and Astrophysics Research Centre, 
5071 W. Saanich Rd. Victoria, BC, V9E 2E7, Canada}
\affiliation{Department of Physics \& Astronomy, University of Victoria, Finnerty Road, Victoria, BC, V8P 1A1, Canada}

\author{Luca Cortese}\affiliation{International Centre for Radio Astronomy Research, The University of Western Australia, 35 Stirling Hwy, 6009 Crawley, WA, Australia }
\affiliation{ARC Centre of Excellence for All Sky Astrophysics in 3 Dimensions (ASTRO 3D), Australia}

\author{Barbara Catinella}\affiliation{International Centre for Radio Astronomy Research, The University of Western Australia, 35 Stirling Hwy, 6009 Crawley, WA, Australia }
\affiliation{ARC Centre of Excellence for All Sky Astrophysics in 3 Dimensions (ASTRO 3D), Australia}

\author[0000-0001-9557-5648]{A. Fraser-McKelvie}\affiliation{European Southern Observatory, Karl-Schwarzschild-Stra{\ss}e 2, Garching, 85748, Germany}

\author{Adam B. Watts}\affiliation{International Centre for Radio Astronomy Research, The University of Western Australia, 35 Stirling Hwy, 6009 Crawley, WA, Australia}

\author[0000-0002-8553-1964]{Amirnezam Amiri}
\affiliation{School of Astronomy, Institute for research in fundamental sciences
(IPM), Tehran, P.O. Box 19395-5531, Iran}
\affiliation{Department of Physics, University of Arkansas, 226 Physics Building, 825 West Dickson Street, Fayetteville, AR 72701, USA}

\author{Alessandro Boselli}\affiliation{Aix-Marseille Universit\'{e}, CNRS, CNES, LAM, Marseille, France}

\author{Woorak Choi}\affiliation{Department of Physics and Astronomy, McMaster University, 1280 Main Street West, Hamilton, Ontario L8S 4M1, Canada}

\author{Aeree Chung}\affiliation{Department of Astronomy, Yonsei University, 50 Yonsei-ro, Seodaemun-gu, Seoul 03722, South Korea}

\author[0000-0003-4932-9379]{Timothy A. Davis}\affiliation{Cardiff Hub for Astrophysics Research \&\ Technology, School of Physics \&\ Astronomy, Cardiff University, Queens Buildings, Cardiff, CF24 3AA, UK}

\author{Eric Emsellem}\affiliation{European Southern Observatory, Karl-Schwarzschild-Stra{\ss}e 2, Garching, 85748, Germany}

\author[0000-0002-1640-5657]{Pavel J\'achym}
\affiliation{Astronomical Institute of the Czech Academy of Sciences, Bo\v cn\'i II 1401, 141 00, Prague, Czech Republic}

\author[0000-0002-9165-8080]{Mar\'ia J. Jim\'enez-Donaire}
\affiliation{AURA for the European Space Agency (ESA), Space Telescope Science Institute, 3700 San Martin Drive, Baltimore, MD 21218 USA}
\affiliation{Observatorio Astron\'omico Nacional (IGN), Alfonso XII 3, 28014, Madrid, Spain}

\author[0000-0002-3365-9210]{Tutku Kolcu}
\affiliation{School of Physics and Astronomy, University of Nottingham, University Park, Nottingham NG7 2RD, UK}

\author[0000-0002-3810-1806]{Bumhyun Lee}
\affiliation{Department of Astronomy, Yonsei University, 50 Yonsei-ro, Seodaemun-gu, Seoul 03722, Republic of Korea}

\author[0000-0003-2723-0810]{Andrei Ristea}
\affiliation{Centre for Astrophysics and Supercomputing, Swinburne University of Technology, Hawthorn, Victoria 3122, Australia}
\affiliation{ARC Centre of Excellence in Optical Microcombs for Breakthrough Science (COMBS)}

 \author[0000-0003-2552-0021]{Jesse van de Sande}
\affiliation{School of Physics, University of New South Wales, NSW 2052, Australia}

\author[0000-0002-0956-7949]{Kristine Spekkens}\affiliation{Department of Physics, Engineering Physics, and Astronomy, Queen’s University, Kingston, ON K7L 3N6, Canada}

 \author[0000-0003-1820-2041]{Sabine Thater}
\affiliation{Department of Astrophysics, University of Vienna, T\"urkenschanzstra\ss e 17, 1180 Vienna}

\author{Christine D. Wilson}\affiliation{Department of Physics and Astronomy, McMaster University, 1280 Main Street West, Hamilton, Ontario L8S 4M1, Canada}

\author[0000-0001-7732-5338]{Nikki Zabel}\affiliation{Department of Astronomy, University of Cape Town, Private Bag X3, Rondebosch 7701, South Africa}

\newcommand{\UOA}
{Department of Physics, University of Arkansas, 226 Physics Building, 825 West Dickson Street, Fayetteville, AR 72701, USA}

\begin{abstract}
We present early science results from the MAUVE (Multiphase Astrophysics to Unveil the Virgo Environment) program which targets 40 Virgo Cluster galaxies to investigate the effect of environment on the interstellar medium (ISM) at $\sim100$~pc scales. From 12 galaxies in the MAUVE--MUSE early sample, we find systematically elevated line ratios compared to PHANGS--MUSE field disks, with higher medians of [\NII]/\Ha\ (0.75 vs.\ 0.50), [\SII]/\Ha\ (0.57 vs.\ 0.49), and [\OIII]/\Hb\ (1.04 vs.\ 0.68). Spatially resolved BPT diagrams show $74$\% of MAUVE--MUSE spaxels ionized by sources other than \HII~regions, versus $61$\% in the field, and we find these ionization differences to be closely coupled to broadened kinematics. 44\% of MAUVE--MUSE spaxels exceed \Ha~$\sigma_{\rm LOS}=40$ km s$^{-1}$ (vs.\ 26\% in the field), driven mainly by non–star-forming gas with $\sigma_{\rm LOS}$ between 40 and 80 km s$^{-1}$, consistent with enhanced contribution of diffuse ionized gas (DIG). A subdominant tail of 5\% of spaxels at $\sigma_{\rm LOS}>100$ km s$^{-1}$, largely absent in PHANGS--MUSE (1\%), points to shocks or turbulent mixing layers from intracluster interactions. Our results show that environmental quenching primarily suppresses star formation, unveiling DIG as the dominant ionized component in cluster disks.  The elevated line ratios and broadened kinematics observed in the MAUVE sample reflect the physical state of the ISM in the absence of vigorous star formation, rather than widespread direct environmental excitation. The observed shock-like emission provides an additional, secondary contribution likely driven by active interactions with the intracluster medium. \end{abstract}

\section{Introduction} \label{sec:intro}

Galaxy clusters are extreme environments that can dramatically alter the evolutionary trajectories of their member galaxies. Spiral galaxies infalling into a cluster experience interactions with the hot intracluster medium (ICM) and with other cluster members that can strip, heat, or remove their gas, rapidly quenching star formation. Key mechanisms include ram pressure stripping by the ICM headwind \citep{Gunn1972}, starvation by cutting off gas accretion \citep{Cowie1977}, and high-speed gravitational encounters or harassment \citep{Moore1996}. These environmental processes help explain the longstanding observations that cluster populations contain a higher fraction of gas-poor, quiescent galaxies than the field \citep[e.g.,][]{Dressler1980,Haynes1984, Haynes1986,Peng2010, Boselli2014, Brown2017, Cortese2021, Boselli2022}. Direct signatures of environment-driven gas loss are seen in the atomic and molecular gas reservoirs of cluster spirals, which are often significantly depleted (up to $\sim0.5-2$ dex) relative to field galaxies of similar stellar mass \citep{Giovanelli1983,Solanes2001,Boselli2014,Brown2017, Zabel2022}. As a result, cluster disks tend to have suppressed and truncated star formation \citep{Koopmann2004, Boselli2006, Cortese2012}.

Environmental processes can also transform the physical state of the interstellar medium (ISM). Recent integral field spectroscopy (IFS) surveys of `jellyfish' galaxies undergoing extreme ram pressure stripping have revealed that extraplanar gas tails often exhibit widespread shock excitation \citep{Yoshida2012, Fossati2016, Poggianti2017, Pedrini2022}. These observations show strong enhancements in low-ionization line ratios such as [\NII]/\Ha~and [\SII]/\Ha, and sometimes elevated [\OIII]/\Hb, consistent with shocks, turbulent mixing, or diffuse ionized gas (DIG) rather than photoionization by young stars \citep{Allen2008, Rich2011, Zhang2017, Belfiore2022}. One key question is whether this environmentally driven shock excitation plays a widespread role in transforming the ISM of cluster galaxies, or whether it is confined to a subset of galaxies, or localized regions, experiencing the most extreme ram pressure conditions.

Shock excitation and DIG are not unique to galaxy clusters. IFS surveys of representative samples of the local galaxy population show some degree of low-ionization (often termed LINER-like) emission, particularly in the DIG component and outer disks \citep{Singh2013,Erroz-Ferrer2019,denBrok2020, Belfiore2022}. In these regions, the harder ionization field compared to star-forming \HII~ regions is powered by hot low mass evolved stars \citep[HOLMES; e.g.,][]{Binette1994, Flores-Fajardo2011, Belfiore2016}, leaked photons from \HII~regions \citep[e.g.,][]{Ferguson1996, Pellegrini2020}, or low-level shocks \citep{Collins2001, Allen2008, Molina2018}. As a result, modest enhancements in [\NII]/\Ha~and [\SII]/\Ha~are common even in isolated systems. Disentangling the environmental contribution to elevated low-ionization line ratios therefore requires not only careful comparison between cluster and field galaxies, but also the use of high spatial resolution IFS to separate localized star-forming regions from diffuse or shock-excited gas across the disk.

In this context, the two open questions addressed by our work are: 1) {\em how does the cluster environment alter the ionization conditions in the ISM?}, and 2) {\em do cluster galaxies show evidence for widespread environment-driven shock excitation across their disks, not just in stripped tails?}

We leverage new data from the {\em Multiphase Astrophysics to Unveil the Virgo Environment} (MAUVE\footnote{\url{https://mauve.icrar.org/}}) survey, a multiwavelength program targeting 40 Virgo Cluster galaxies \citep{Catinella2025}. MAUVE aims to track environmental effects across ionized and molecular gas phases, as well as the stellar populations, of spiral galaxies during their evolution within the cluster. In particular, we use the initial observations from MAUVE--MUSE, a European Southern Observatory (ESO) Large Program targeting these galaxies with the Multi Unit Spectroscopic Explorer \citep[MUSE;][]{Bacon2010} to obtain optical IFS \citep{Watts2024}. The average seeing during the observations was $\sim$0.85\arcsec (ranging from $\sim$0.5\arcsec\ to 1.4\arcsec), equivalent to $\sim 68$ pc at the distance of Virgo.

We assume a distance of 16.5 Mpc for Virgo Cluster galaxies \citep{Mei2007}, and, where necessary, physical properties are derived assuming $H_{0} = 70~\mathrm{km~s^{-1}~Mpc^{-1}}$, a flat cosmology with $\Omega_{\mathrm{m}} = 0.27$, and a \citet{Kroupa2001} initial mass function (IMF).

This paper is organized as follows. We describe the MAUVE--MUSE sample, observations, and data reduction in Sections \ref{sec:Sample} and \ref{sec:Data}. In Section \ref{sec:Results}, we present a comparative analysis of ionized gas properties between MAUVE--MUSE cluster galaxies and field systems from the PHANGS--MUSE survey \citep{Emsellem2022}--part of the PHANGS (Physics at High Angular resolution in Nearby GalaxieS) project--discussing the implications for environmental transformation of galaxies and their ISM in Section \ref{sec:discussion}. Our conclusions are summarized in Section \ref{sec:conclusions}. Lastly, we provide an atlas of emission line flux and flux ratio maps for the 12 MAUVE--MUSE galaxies used in this work in Appendices \ref{appendix:emlines_atlas} and \ref{appendix:line_ratio_atlas}, respectively.

\section{The MAUVE--MUSE Early Science Sample} \label{sec:Sample}
\begin{figure*}
    \centering
    \includegraphics[width=\textwidth]{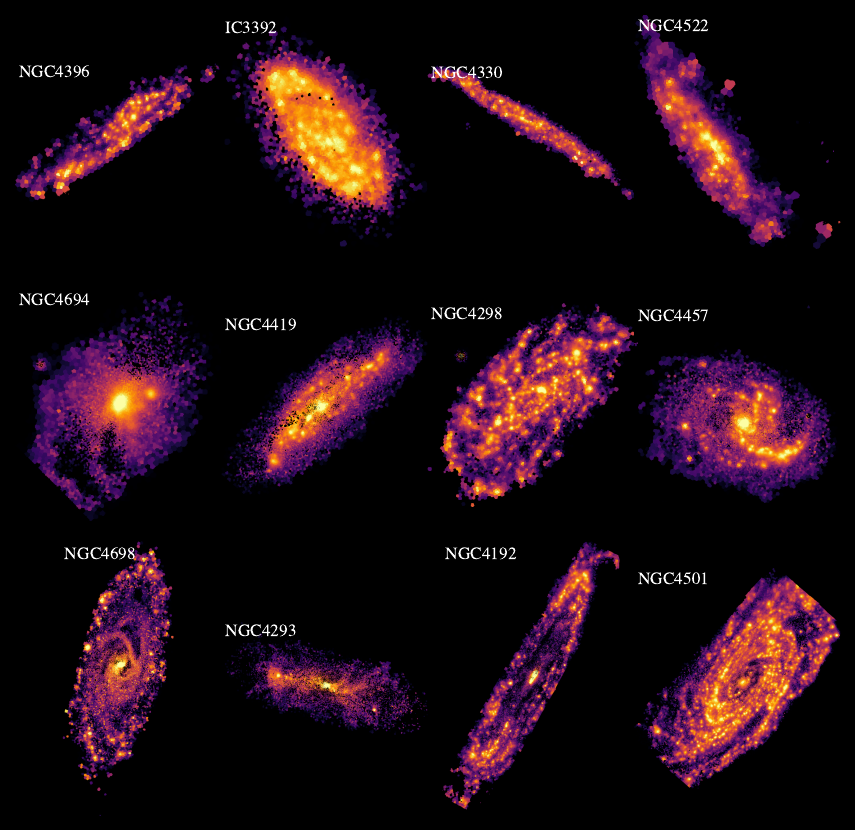}
    \caption{Voronoi-binned \Ha~maps for the MAUVE--MUSE early science sample. Galaxies are ordered by increasing stellar mass given in Table \ref{tab:galaxy_properties}. Once the survey is complete, the final sample will comprise 40 galaxies. Maps are not shown on the same physical scale.}
    \label{fig:ha_atlas}
\end{figure*}

\begin{deluxetable*}{lcccccc}
\tablecaption{Galaxy properties for the MAUVE--MUSE early science sample.}
\tablehead{\colhead{Galaxy} & \colhead{RA (J2000)} & \colhead{Dec (J2000)} & \colhead{Inclination} & \colhead{$\log~M_\star$} & \colhead{$\log~\mathrm{SFR}$} & \colhead{\HI-def}\\ \colhead{ } & \colhead{deg} & \colhead{deg} & \colhead{deg} & \colhead{$\mathrm{M_{\odot}}$} & \colhead{$\mathrm{M_{\odot}\,yr^{-1}}$} & \colhead{ }}
\startdata
IC~3392 & 187.1803 & 14.9993 & 68 & 9.8$\pm$0.1 & -1.0$\pm$0.2 & 1.2$\pm$0.1 \\
NGC~4192 & 183.4524 & 14.8992 & 83 & 10.8$\pm$0.1 & 0.1$\pm$0.2 & 0.5$\pm$0.2 \\
NGC~4293 & 185.3061 & 18.3842 & 67 & 10.5$\pm$0.1 & -0.2$\pm$0.2 & 2.2$\pm$0.2 \\
NGC~4298 & 185.3880 & 14.6055 & 52 & 10.1$\pm$0.1 & -0.3$\pm$0.2 & 0.4$\pm$0.0 \\
NGC~4330 & 185.8206 & 11.3678 & 90 & 9.6$\pm$0.2 & -0.9$\pm$0.2 & 0.8$\pm$0.0 \\
NGC~4396 & 186.4986 & 15.6695 & 83 & 9.3$\pm$0.1 & -0.7$\pm$0.2 & 0.3$\pm$0.0 \\
NGC~4419 & 186.7348 & 15.0476 & 74 & 10.2$\pm$0.1 & 0.1$\pm$0.2 & 1.4$\pm$0.2 \\
NGC~4457 & 187.2459 & 3.5706 & 37 & 10.5$\pm$0.1 & -0.4$\pm$0.2 & 0.9$\pm$0.2 \\
NGC~4501 & 187.9972 & 14.4197 & 65 & 11.0$\pm$0.1 & 0.4$\pm$0.2 & 0.6$\pm$0.1 \\
NGC~4522 & 188.4155 & 9.1741 & 82 & 9.6$\pm$0.1 & -0.8$\pm$0.2 & 0.9$\pm$0.0 \\
NGC~4694 & 192.0628 & 10.9835 & 62 & 9.9$\pm$0.1 & -0.9$\pm$0.2 & 0.8$\pm$0.2 \\
NGC~4698 & 192.0958 & 8.4875 & 66 & 10.5$\pm$0.1 & -0.8$\pm$0.2 & 0.0$\pm$0.2
e\enddata
\tablecomments{\footnotesize Stellar masses and star formation rates are from \citet{Leroy2019}, corrected to our adopted Virgo distance of 16.5~Mpc. $r$-band inclinations are taken from \citet{Brown2021}. \HI~deficiencies are from \citet{Yoon2017}, defined as the logarithmic difference between the expected \HI~mass of an isolated galaxy of the same morphological type and diameter, and the observed \HI~mass.}
\label{tab:galaxy_properties}
\end{deluxetable*}

Our analysis sample comprises 12 MAUVE--MUSE galaxies fully observed and reduced as of December 2024 (see below for sample selection). The galaxies span a broad range of galaxy properties: stellar masses $\log(M_{\star}/M_{\odot}) = 9.4 - 11.0$, star formation rates $SFR=0.1-2.7 \, \mathrm{M_{\odot}\,yr}^{-1}$, inclinations $i=37^{\circ}-90^{\circ}$ with a median $i=66^{\circ}$, and~\HI~deficiencies from $0.0$ to $2.3$~dex, capturing systems from HI-normal to strongly gas-depleted where positive values indicate gas depletion relative to the field control sample (see Table \ref{tab:galaxy_properties}). This diversity enables a detailed investigation of the environmental effects on galaxy gas content, ionization conditions, and star formation activity across a range of evolutionary stages. 
% A thorough description of the full survey sample selection and data reduction will be provided in Cortese et al. (in prep).

Our sample also spans a wide range of cluster infall stages, morphological disturbances, and ionization conditions \citep{Koopmann2004, Chung2009, Yoon2017, Brown2017}. The resulting diversity can be seen in the illustrative atlas of \Ha\ flux maps shown in Figure \ref{fig:ha_atlas}. Several systems are in early stages of infall, clearly pre-pericenter, with little or no evidence of strong ram-pressure effects within the disc, though in some cases with clear \HI\ tails \citep[NGC~4192, NGC~4298, NGC~4396, NGC~4501;][]{Cayatte1990, Chung2009, Vollmer2008a, Vollmer2013, Nehlig2016}. Archetypal examples of galaxies experiencing peak ram-pressure stripping are also included \citep[NGC~4330, NGC~4522;][]{Kenney1999, Abramson2011, Vollmer2013, Fossati2018, Vollmer2021}.  
IC~3392 is post-pericenter and characterized by truncated gas disc \citep[IC~3392;][]{Koopmann2001, Chung2009}. Finally, a subset of galaxies show more complex evolutionary paths. Many of these are \HI-deficient, suggesting they have already passed through the cluster, but their present-day properties imply that additional environmental mechanisms beyond ram pressure have influenced their evolution. Clear examples include NGC~4293, NGC~4694, NGC~4419, NGC~4457, and NGC~4698 \citep{vanDriel1989, Bertola1999, Chung2009, Vollmer2013}. 

% Several systems are in pre-pericentre or early-infall phases, showing clear or emerging signs of ram pressure stripping and extraplanar gas \citep[NGC~4192, NGC~4298, NGC~4330,NGC~4501, NGC~4522;][]{Cayatte1990, Kenney1999, Vollmer2008a, Vollmer2013, Chung2009}. Others represent post-pericentre or backsplash populations \citep[IC~3392, NGC~4064, NGC~4293, NGC~4419;][]{Kenney1990, Koopmann2001, Cortes2006, Chung2009, Abramson2011, Fossati2018}, characterized by truncated gas disks or enhanced outflows. A subset of galaxies display complex or ambiguous environmental features, such as NGC~4457 and NGC~4694, which show evidence for gravitational interactions or mixed stripping mechanisms \citep{vanDriel1989, Cortes2006, Vollmer2013}, or NGC~4698 which has a quiescent stellar population with a kinematically decoupled bulge and unusually large gas reservoir \citep{Bertola1999, Chung2009, Cortese2009}. NGC~4383 \citep{Watts2024} and NGC~4064 host prominent star formation-driven outflows. To our knowledge, the outflow in NGC~4064 is previously unreported (Attwater et al., subm.).  
% The sample captures a wide range of stripping stages, ionization sources, and interaction signatures, making it well-suited for dissecting the multi-phase impact of environment on galaxy evolution within the Virgo cluster.
\begin{figure}
    \centering
    \includegraphics{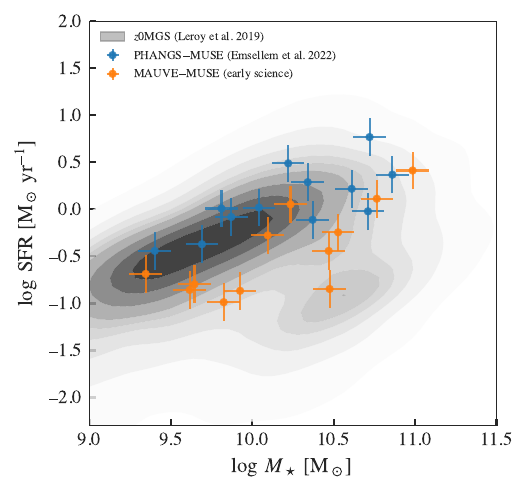}
    \caption{Stellar mass vs.\ SFR for the MAUVE--MUSE early science sample (orange) and PHANGS--MUSE field control galaxies used in this work  (blue). The gray density contours show density distribution of $15,747$ local galaxies from \citet{Leroy2019}.}
    \label{fig:global_main_sequence}
\end{figure}

For our analysis we exclude NGC~4064 and NGC~4383, whose observed ISM properties are strongly impacted by star formation-driven outflows rather than environmental processes. These systems, studied in detail by \citet{Watts2024} and \citet{Attwater2025}, are of intrinsic interest but would complicate our controlled comparison between field and cluster galaxies. To place our analysis sample in context, Figure \ref{fig:global_main_sequence} shows the distribution of MAUVE--MUSE and PHANGS--MUSE galaxies (see Section \ref{sec:phangs}) on the star formation rate–stellar mass (SFR–$M_\star$) plane. The MAUVE--MUSE galaxies span a wide range of SFRs at fixed stellar mass, with many lying below the $z \sim 0$ main sequence. This diversity reflects the heterogeneous impact of environmental processes, ranging from ongoing star formation to various stages of quenching, within the Virgo Cluster. The final MAUVE--MUSE analysis sample used in this paper comprises 12 galaxies with line fluxes detected at $>5\sigma$ in 2.4 million 0.2\arcsec\ spaxels (depending on the emission line) across 73,000 unique Voronoi bins (see Section \ref{sec:Data}).

\subsection{The PHANGS--MUSE Sample}\label{sec:phangs}
Our comparison sample outside of the cluster environment is drawn from the publicly available PHANGS--MUSE survey, which is well matched to the MAUVE--MUSE early science sample in stellar mass, morphology, and sensitivity. The dataset comprises 19 nearby star-forming late-type galaxies at distances ranging from approximately 9 to 19~Mpc. Selected for their relatively low inclinations (typically $i \lesssim 55^\circ$), these galaxies lie close to the star-formation main sequence, covering stellar masses between $\log(M_{\star}/M_{\odot}) \approx 9.4-11.0$ and star formation rates spanning $0.3-16.9~M_{\odot}\,\mathrm{yr}^{-1}$. 
% The MUSE IFU observations predominantly sample the star-forming disk regions, reaching out to an average galactocentric radius of approximately $0.86\,R_{25}$.
% For consistency with our analysis, we exclude galaxies hosting strong AGN activity (NGC~1365, NGC~1512, NGC~1566, NGC~1672) and galaxies located in the Virgo cluster outskirts (NGC~4254, NGC~4321, NGC~4535), which are also included as archival data in the MAUVE--MUSE sample. 
To ensure a representative control sample of main-sequence field galaxies, we exclude systems hosting strong active galacic nuclei (AGN; NGC~1365, NGC~1512, NGC~1566, NGC~1672) as well as galaxies located in the Virgo cluster (NGC~4254, NGC~4321, NGC~4535).
The final control sample thus comprises 12 field galaxies. Full details of galaxy properties and data reduction methods for PHANGS--MUSE are available in \citet{Emsellem2022} and \citet{Santoro2022}.

% Figure~\ref{fig:global_main_sequence} shows the star-forming main sequence for the PHANGS--MUSE and MAUVE--MUSE samples. The MAUVE--MUSE galaxies span a similar stellar mass range as PHANGS--MUSE but are systematically offset to lower star formation rates, lying predominantly below the main sequence.

\section{Data Description} \label{sec:Data}
The MAUVE--MUSE observing strategy is described in \citet{Catinella2025} and will be detailed fully in Cortese et al. in prep, along with the data reduction process. Briefly, all galaxies in our sample were observed as part of the MAUVE VLT/MUSE Large Programme (MAUVE--MUSE, ID~110.244E), using MUSE in its Wide Field Mode. This configuration provides a $1\arcmin \times 1\arcmin$ field of view, $0.2\arcsec$ spaxels, and a spectral sampling of 1.25\,\AA\,pix$^{-1}$ with a typical spectral resolution of $\mathrm{FWHM} \approx 2.5$\,\AA\ at 7000\,\AA. To ensure full coverage of the molecular gas disc traced by ALMA/VERTICO, each galaxy was mapped with between one and nine individual MUSE pointings (four on average), for a total of 142 pointings in the full programme \citep{Catinella2025}. Each pointing consists of four 750\,s on-source exposures (3000\,s total), plus an offset sky exposure, resulting in a total observing-block duration of $\sim 73$\,min. The exposure depth was chosen to achieve ${\rm S/N} \ge 20$ in the stellar continuum at the outskirts of the H$_2$ disc and to detect diffuse ionised gas at ${\rm rms} \sim 10^{-18}$\,erg\,s$^{-1}$\,cm$^{-2}$\,\AA$^{-1}$ on $\sim 200$\,pc scales, requirements that were verified using the pilot observations of NGC~4383. Observations were carried out in Service Mode under clear conditions with a relaxed seeing constraint of $\le 1.3\arcsec$, corresponding to a spatial resolution of $\sim 100$\,pc at the Virgo distance (16.5\,Mpc).

The method for producing MAUVE--MUSE data cubes is described in full in \citet{Watts2024} and is based upon the approach used by the PHANGS--MUSE team and described in \citet{Emsellem2022}. As such, we are able to make direct use of the publicly released PHANGS--MUSE science-ready data cubes from their first data release\footnote{\url{https://www.eso.org/rm/api/v1/public/releaseDescriptions/184}}.

% In the spectral fitting and emission line analysis we target S/N = 40 in the stellar continuum across spatial bins to allow for robust continuum subtraction from the emission-line products presented in this paper. In practice, the difference relative to the PHANGS--MUSE target of ${\rm S/N}=35$ is small and has a negligible effect on the resulting bin sizes. For the PHANGS comparison sample, we do not re-reduce the raw data. Instead, }

In brief, the pipeline builds on the ESO MUSE data processing pipeline \citep{Weilbacher2020}, applying additional steps to ensure science-ready products suitable for spatially resolved emission-line analysis.
This includes improved sky subtraction, accurate astrometric and flux calibration, propagation of variances, and the creation of mosaics from multiple pointings. The final data cubes are corrected for Galactic extinction and registered onto a common astrometric grid, providing homogeneous, high-quality data cubes suitable for both continuum and emission-line studies. For this release we used v2.28.2 of {\tt pymusepipe}\footnote{\url{https://github.com/emsellem/pymusepipe} \citep{Emsellem2022}}.

Emission-line and kinematic maps for the MAUVE--MUSE and PHANGS-MUSE data in this work are produced using the penalized pixel-fitting method 
\citep[\texttt{pPXF};][]{Cappellari2004, Cappellari2017, Cappellari2023} as implemented in the new Galaxy IFU Spectroscopy Tool pipeline \citep[\texttt{nGIST}\footnote{\url{https://geckos-survey.github.io/gist-documentation}};][]{Fraser-McKelvie2025}. 
The \texttt{nGIST} emission-line module builds directly on approaches developed for the MaNGA \citep[Mapping Nearby Galaxies at Apache Point Observatory;][]{Westfall2019} and PHANGS data analysis pipelines \citep{Emsellem2022}, and employs the MILES stellar population models  \citep{Vazdekis2010} to model the stellar continuum. We direct the reader to \citet{Fraser-McKelvie2024} for a complete description of the pipeline implementation.

\begin{figure}
    \centering
    \includegraphics{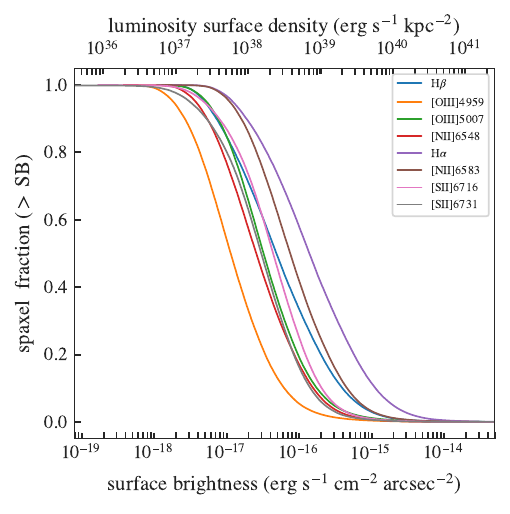}
    \caption{Fraction of detected ($\geq5\sigma$) spaxels greater than a given surface brightness for each emission line in the MAUVE--MUSE sample used in this study. We show the luminosity surface density on the top x-axis. Note that the [\OIII]$\lambda4958$ and [\NII]$\lambda6548$ fluxes are tied to their corresponding doublet line and are included for completeness.
}
    \label{fig:mauve-muse_sb_ecdf}
\end{figure}

Briefly, spaxels with continuum S/N$< 1.5$ over the 4800--7000\,\AA\ range are excluded from the fitting. The stellar continuum is binned to a target S/N$= 40$ using Voronoi spatial binning, after which we perform simultaneous fitting of both stellar continuum and emission lines with \texttt{pPXF}. As is standard for \texttt{pPXF}, fluxes of the [O\,\textsc{iii}]$\lambda4958$ and [N\,\textsc{ii}]$\lambda6548$ lines are tied to their corresponding doublet components. For emission-line moment maps in both MAUVE--MUSE and PHANGS--MUSE, we further mask bins where flux/error $< 3$, a proxy for goodness of fit, or surface brightness $< 5\times10^{-20}$~\text{erg}~\text{s}$^{-1}$~\text{cm}$^{-2}$. This choice was reduces artificial noise and systematics in the maps, confirmed by visual inspection, and provides a clearer definition of the DIG regime probed at the sensitivity limits of our data, without affecting our primary scientific conclusions. The exact values for these masking choices does not affect our conclusions.  All emission lines are fit with single Gaussians; we note that this can artificially broaden measured velocity dispersions in regions with multiple unresolved kinematic components. The velocity dispersions are corrected for the MUSE instrumental broadening following \citet{Bacon2010}. We correct the line fluxes for dust attenuation using the Balmer decrement (assuming the \Ha/\Hb~ flux ratio = 2.86, appropriate for case B recombination at $T_e = 10^4$ K and $n_e = 100$ cm$^{-3}$). To prevent unphysical extinction corrections, any observed ratio below the theoretical expectation is set to 2.86. The color excess $E(B-V)$ is calculated for each bin using the \citet{Cardelli1989} extinction law, parameterized with $R_V = 3.1$, and extinction coefficients $k(\lambda)$ computed at the wavelengths of \Ha~ and \Hb. From this, we derive the $V$-band extinction $A_V$, and correct all emission line fluxes accordingly. For each emission line, the attenuation $A(\lambda)$ is computed based on the line's central wavelength, and fluxes are corrected using the same \citet{Cardelli1989} extinction law.

Emission-line fluxes, line ratios, and kinematics are measured from spectra that have been Voronoi binned according to the target continuum signal-to-noise described above. The resulting best-fitting quantities are assigned to each spaxel belonging to the corresponding bin, producing maps sampled on the native $0.2^{\prime\prime}$ spaxel grid but with piecewise-constant values within each bin. These spaxel-level distributions can therefore be interpreted as area-weighted summaries of the binned maps rather than as samples of statistically independent measurements. All quantitative distributions are computed over these spaxels, with per-spaxel weights $w_i = 1/(N_{spax,\;gal(i)}\,N_{gal})$ so that each galaxy contributes equally to the combined distributions irrespective of its number of valid spaxels.

Figure \ref{fig:mauve-muse_sb_ecdf} shows the fraction of MAUVE--MUSE sample spaxels above a given surface brightness or luminosity surface density for each emission line used in this work: \Hb, [\OIII]~$\lambda4959$ and $\lambda5007$, [\NII]~$\lambda6548$ and $\lambda6583$, \Ha, and [\SII]~$\lambda6716$ and $\lambda6731$. The \Ha~surface brightness is typically (95\%) greater than $1.2 \times 10^{-17}$ erg s$^{-1}$ cm$^{-2}$ arcsec$^{-2}$, equivalent to $4.8 \times 10^{-19}$ erg s$^{-1}$ cm$^{-2}$ spaxel$^{-1}$. This corresponds to a luminosity surface density of $6.2 \times 10^{37}$ erg s$^{-1}$ kpc$^{-2}$. For our faintest line, [\OIII]4959, spaxels are typically brighter than $1.8 \times 10^{-18}$ erg s$^{-1}$ cm$^{-2}$ arcsec$^{-2} = 7.3 \times 10^{-20}$ erg s$^{-1}$ cm$^{-2}$ spaxel$^{-1}$, and luminosity surface density of $9.3 \times 10^{36}$ erg s$^{-1}$ kpc$^{-2}$.

\begin{figure*}
    \centering
    \includegraphics{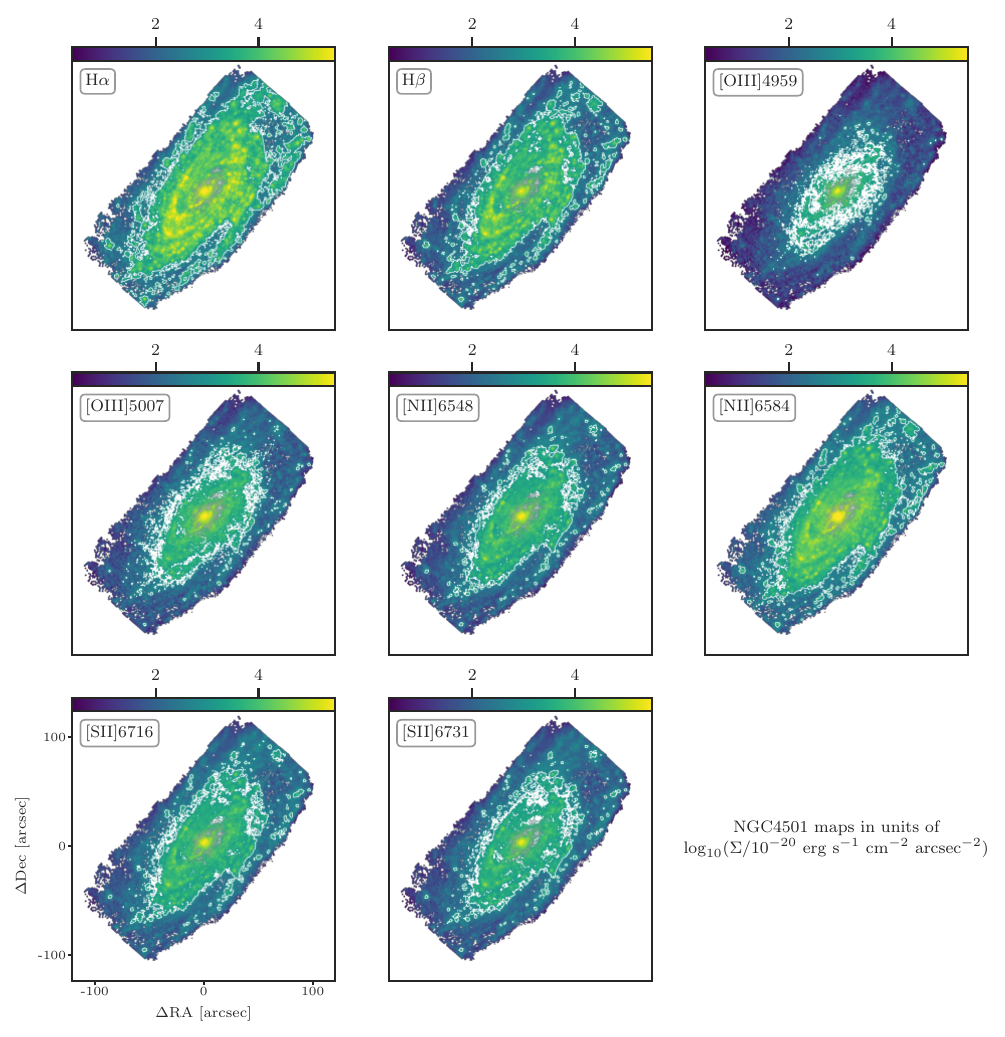}
    \caption{Emission line maps for NGC~4501, an early-stage ram pressure stripped galaxy in the MAUVE--MUSE sample. Shown are the extinction-corrected fluxes of H$\alpha$, H$\beta$, [\OIII]$\lambda\lambda$4958,5006, [\NII]$\lambda\lambda$6548,6583, and [\SII]$\lambda\lambda$6716,6730, masked where $\mathrm{flux/error} < 3$ and surface brightness $< 5\times 10^{-20}$~\text{erg}~\text{s}$^{-1}$~\text{cm}$^{-2}$~\text{arcsec}$^{-2}$. The thin white contour denotes $10^{-17}$~\text{erg}~\text{s}$^{-1}$~\text{cm}$^{-2}$~\text{arcsec}$^{-2}$2.} Fluxes have been corrected for dust attenuation using the Balmer decrement. [\OIII]4958 and [\NII]6548 fluxes are derived from their doublet line ratios and included for completeness. 
    Emission line maps for all MAUVE--MUSE galaxies used in this work are available in Appendix~\ref{appendix:emlines_atlas}.
    \label{fig:emlines_ngc4501}
\end{figure*}

\begin{figure*}
    \centering
    \includegraphics{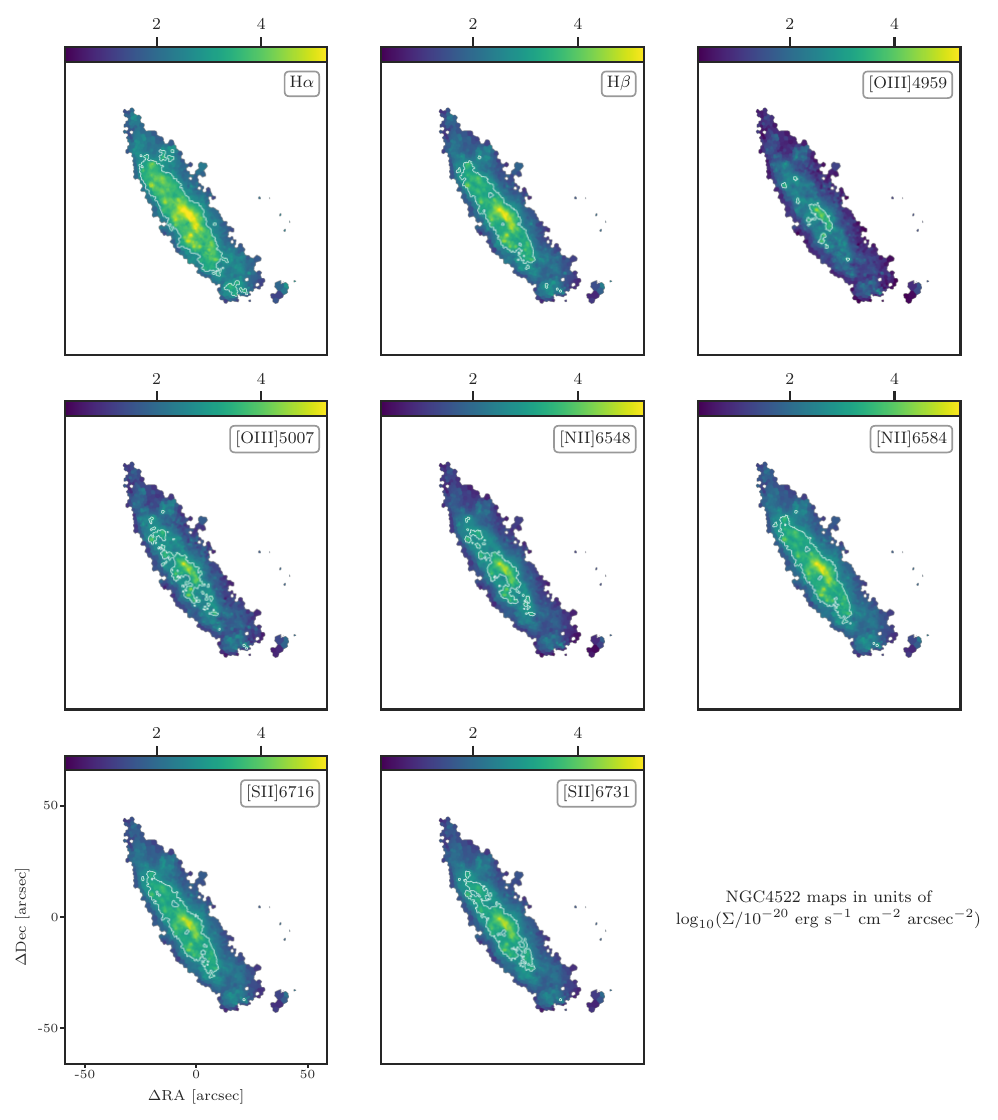}
    \caption{Emission line maps for NGC~4522, a strongly stripped Virgo spiral with prominent extraplanar ionized gas. Panels are the same as Figure \ref{fig:emlines_ngc4501}. Emission line maps for all MAUVE--MUSE galaxies used in this work are available in Appendix~\ref{appendix:emlines_atlas}.}
    \label{fig:emlines_ngc4522}
\end{figure*}

% \subsection{Emission Line Fluxes}
Figures \ref{fig:emlines_ngc4501} and \ref{fig:emlines_ngc4522} show the MAUVE--MUSE emission line surface brightness maps for two MAUVE galaxies that we later analyze as ram pressure stripping case studies, NGC~4501 and NGC~4522. The data shown are available for all targets and the full collection of MUSE optical emission line maps for our sample can be found in Appendix \ref{appendix:emlines_atlas}. Each flux map is in units of $\log$($10^{-20}$~\text{erg}~\text{s}$^{-1}$~\text{cm}$^{-2}$~\text{arcsec}$^{-2}$). NGC 4501 is a pre-pericenter spiral galaxy exhibiting early signs of environmental interaction. The presence of a faint tail in \HI\ and radio continuum emission suggests the onset of ram pressure stripping \citep{Vollmer2008a}. NGC~4522, by contrast, is a lower-mass system in an advanced phase of ram pressure stripping affecting atomic and molecular gas phases \citep{Kenney1999, Vollmer2008b, Lee2018}. It exhibits a well-known extraplanar ISM tail, clearly visible in our emission line maps which show the ionized material being stripped from the disk.

\section{Results} \label{sec:Results}
In this section, we provide an analysis of the impact of the Virgo Cluster environment on the physical conditions within the ionized ISM.

\subsection{Emission Line Ratio and Kinematic Maps}
\label{sec:emission_line_ratios}
\begin{figure*}
    \centering
    \includegraphics{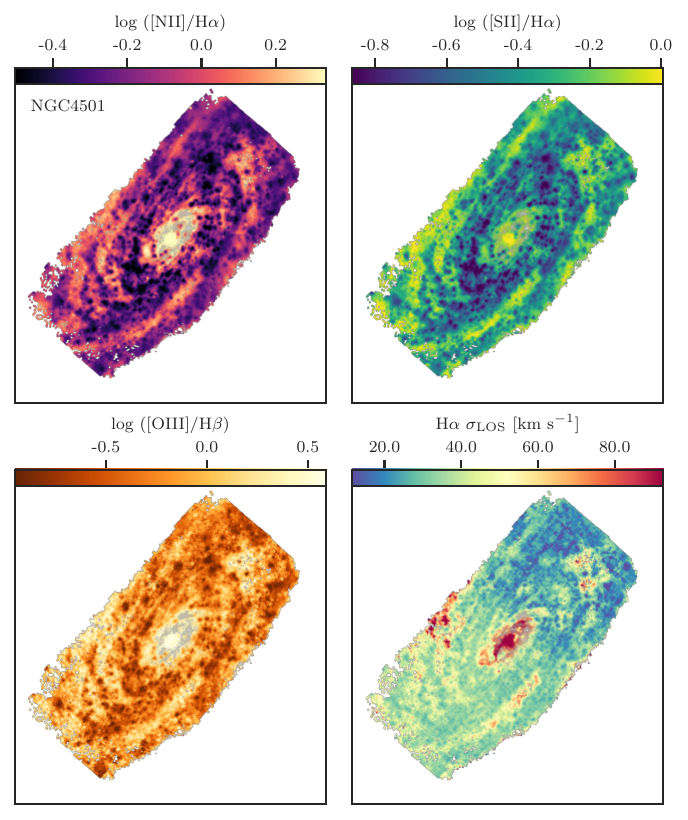}
    \caption{Emission line ratio maps for NGC~4501. Panels (left to right, top to bottom) show the [\NII]/\Ha, [\SII]/\Ha, and [\OIII]/\Hb, and  \Ha~ line widths along the line-of-sight (\Ha~$\sigma_{\rm LOS}$). The full set of MAUVE--MUSE emission line ratio maps is provided in Appendix \ref{appendix:line_ratio_atlas}.}
    \label{fig:NGC4501_emline_ratios}
\end{figure*}

\begin{figure*}
    \centering
    \includegraphics{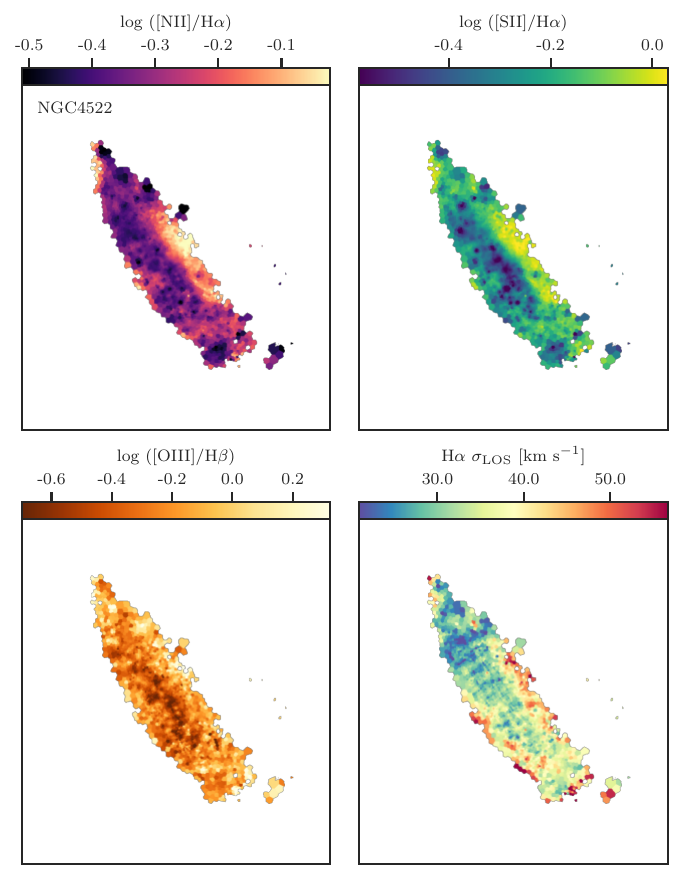}
    \caption{Same as Figure \ref{fig:NGC4501_emline_ratios} for NGC~4522.}
    \label{fig:NGC4522_emline_ratios}
\end{figure*}

Figures \ref{fig:NGC4501_emline_ratios} and \ref{fig:NGC4522_emline_ratios} show key line ratios ([\NII]/\Ha, [\SII]/\Ha, [\OIII]/\Hb) and kinematics for the two example galaxies. Figure~\ref{fig:NGC4501_emline_ratios} shows that the ionized gas in NGC~4501 is primarily confined to the disk with an enhancement of [\NII]/\Ha\ and [\SII]/\Ha\ observed in the central regions. The increased ratios are more pronounced in [\NII]/\Ha, consistent with ionization from evolved stellar populations and/or increased metallicity, especially in the center. Notably, a large region of high [\NII]/\Ha\ and [\SII]/\Ha\ is also observed in the eastern disk, spatially coincident with the nascent~\HI~and radio continuum tail \citep{Chung2009, Edler2023}. This suggests that ram pressure stripping, even in an early phase, is already imprinting on the excitation conditions of the ISM within the disk. Indeed, while the [\OIII]/\Hb\ ratio remains low across much of the galaxy, indicating a low ionization parameter, the highest [\OIII]/\Hb\ values are also found in the east region. These high line ratios are consistent with shock excitation and/or a DIG component that is powered by either \HII\ leakage or evolved stellar populations \citep[e.g.,][]{Rich2011, Belfiore2022}. The \Ha~$\sigma_{\rm LOS}$ maps show localized enhancements in the east and interarm regions coincident with the stripping of NGC~4501, whereas in NGC~4522 (Fig. \ref{fig:NGC4522_emline_ratios}) the line widths are systematically elevated along the leading-edge (south-east) and extraplanar regions compared to the disk, reflecting the stronger turbulence associated with gas that is being actively compressed and/or stripped.

In NGC~4522, a classical case of ram pressure stripping, the Balmer decrement is elevated above 2.86 across the stripped tail, indicating the presence of dust and/or enhanced attenuation associated with stripped material, in agreement with recent observations \citep{Longobardi2020}. Elevated [\NII]/\Ha\ and [\SII]/\Ha\ ratios in the extraplanar regions suggest a significant contribution from shocks and/or stripping of the DIG from the disk. The [\OIII]/\Hb\ ratio shows localized enhancements near the extraplanar regions, potentially tracing areas of compression or turbulent mixing.

\subsection{Emission Line Ratio Distributions}

\begin{figure*}
    \centering
    \includegraphics{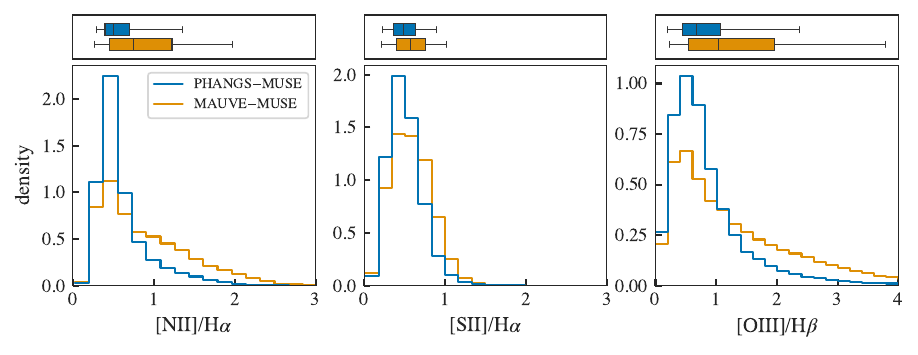}
    \caption{Distributions of [\NII]/\Ha, [\SII]/\Ha, and [\OIII]/\Hb\ for PHANGS--MUSE (blue) and MAUVE--MUSE (orange) spaxels. Top panels: box plots showing the weighted percentiles (P$_5$, P$_{25}$, Median, P$_{75}$, P$_{95}$) listed in Table~\ref{tab:line_ratio_stats}. Bottom panels: weighted histograms showing the underlying distribution shapes. Compared to the field galaxies, cluster galaxies exhibit systematically elevated line ratios. Spaxels are weighted such that each galaxy contributes equally to the distribution.}
    \label{fig:line_ratios_phangs_mauve}
\end{figure*}

\begin{deluxetable*}{llcccccc}
\tablecaption{Weighted line ratio (P$_5$, P$_{25}$, Median, P$_{75}$, P$_{95}$) and KDE peaks for PHANGS--MUSE and MAUVE--MUSE samples. The percentiles are shown in the box plots of Figure \ref{fig:line_ratios_phangs_mauve}.\label{tab:line_ratio_stats}}
\tablehead{
\colhead{Survey} & \colhead{Line Ratio} & \colhead{P$_5$} & \colhead{P$_{25}$} & \colhead{Median} & \colhead{P$_{75}$} & \colhead{P$_{95}$} & \colhead{KDE peak}
}
\startdata
PHANGS--MUSE & {}[\NII]/H$\alpha$  & 0.29 & 0.40 & 0.50 & 0.70 & 1.36 & 0.41 \\
             & {}[\SII]/H$\alpha$  & 0.23 & 0.36 & 0.49 & 0.63 & 0.89 & 0.44 \\
             & {}[\OIII]/H$\beta$  & 0.21 & 0.44 & 0.68 & 1.08 & 2.37 & 0.55 \\
\hline
MAUVE--MUSE  & {}[\NII]/H$\alpha$  & 0.27 & 0.46 & 0.75 & 1.22 & 1.97 & 0.41 \\
             & {}[\SII]/H$\alpha$  & 0.22 & 0.40 & 0.57 & 0.75 & 1.02 & 0.45 \\
             & {}[\OIII]/H$\beta$  & 0.23 & 0.54 & 1.04 & 1.96 & 3.79 & 0.44 \\
\enddata
\tablecomments{Spaxels are weighted such that each galaxy contributes equally to the distribution of each survey. KDE peak is defined as the mode of the distribution, computed using a Gaussian kernel density estimate from \texttt{scipy.stats.gaussian\_kde}, evaluated over a fine grid of the observed range.}
\end{deluxetable*}

To extend our analysis of emission line ratios across the full sample, Figure~\ref{fig:line_ratios_phangs_mauve} compares the distributions of [\NII]/\Ha, [\SII]/\Ha, and [\OIII]/\Hb\ for PHANGS--MUSE and MAUVE--MUSE spaxels. Each distribution is weighted (see Section \ref{sec:Data}) ensure equal contribution from each galaxy and normalized to account for differing spaxel counts. The box plots highlight the weighted percentiles listed in Table~\ref{tab:line_ratio_stats}, while the histograms show the underlying distribution shapes. 

The galaxies in MAUVE–MUSE exhibit systematically elevated emission line ratios compared to field galaxies, consistent with an increased contribution from non–H \textsc{ii} region ionization processes such as shocks, diffuse ionized gas powered by evolved stellar populations, or leakage of photons from \HII\ regions. For [\NII]/\Ha, the median increases from 0.50 (PHANGS–MUSE) to 0.75 (MAUVE–MUSE), and the interquartile range (IQR = P$_{75}$–P$_{25}$) broadens from 0.30 to 0.76, with the 90\% range (P$_{95}$–P$_{5}$) expanding from 1.07 to 1.70. [\SII]/\Ha\ shows a comparable shift, with the median increasing from 0.49 to 0.57 and the percentile ranges also broadening (IQR: 0.27 to 0.35; 90\% range: 0.66 to 0.80). The [\OIII]/\Hb~median rises from 0.68 to 1.04 and the 90\% range increases from 2.16 to 3.56. These systematic enhancements reflect a shift in the balance of ionized gas toward the diffuse component as star formation is suppressed, demonstrating that cluster disks are dominated by an ionization environment more diverse and harder than that of typical field spirals.

\subsection{BPT Diagnostic Diagrams}
\begin{figure*}
    \centering
    \includegraphics{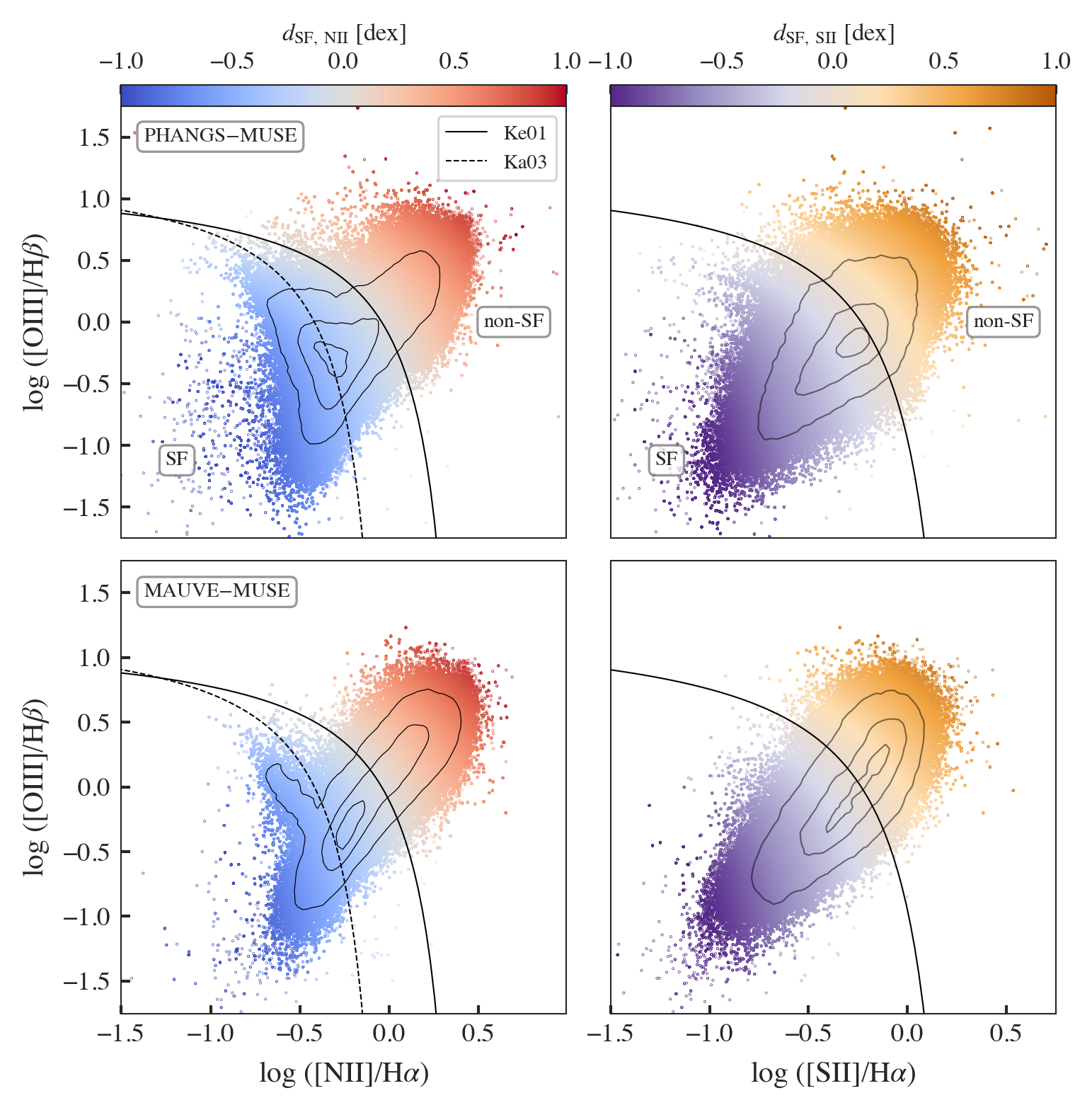}
    \caption{
    Spatially resolved BPT diagrams for the PHANGS--MUSE (top) and MAUVE--MUSE (bottom) samples. 
    Left: [\NII]/\Ha\ vs.\ [\OIII]/\Hb. Right: [\SII]/\Ha\ vs.\ [\OIII]/\Hb. Colors indicate the orthogonal distance, $d_{\mathrm{SF}}$, in dex from the empirical Ka03 (dashed) and theoretical Ke01 (solid) star-formation boundaries, where positive values denote increasing departure from pure star-formation excitation toward non--H\,\textsc{ii} region regimes. Contours mark the 10th, 50th, and 90th percentiles of spaxel density. MAUVE--MUSE galaxies show a clear excess of spaxels at large positive $d_{\mathrm{SF}}$, indicating widespread departures from star-formation excitation compared to the field.
    }
    \label{fig:sf_distance_bpt}
\end{figure*}
The enhanced line ratios observed in the MAUVE--MUSE sample raise the question of whether the ionization conditions in cluster galaxies are systematically different from those in the field. To address this, we turn to the spatially resolved Baldwin-Phillips-Terlevich (BPT) diagrams \citep{Baldwin1981}, which allow us to distinguish between gas excited by star formation and that influenced by harder sources such as shocks or emission from older stellar populations \citep{Veilleux1987,Kewley2001,Kauffmann2003, Allen2008, Belfiore2022}.

Figure~\ref{fig:sf_distance_bpt} shows the distributions of spaxel-level line ratios in PHANGS--MUSE (field) and MAUVE--MUSE (cluster) galaxies, for both [\NII]/\Ha~and [\SII]/\Ha-based diagnostics. Colors encode the orthogonal distance, $d_{\mathrm{SF}}$, from the empirical \citet[][Ka03, dashed]{Kauffmann2003} star-formation and theoretical \citet[][Ke01, solid]{Kewley2001} starburst boundaries, with positive values denoting increasing departure from pure star-formation excitation toward non--H\,\textsc{ii} region regimes. 
% This definition is conceptually related to other composite diagnostics such as the softness parameter \citep{Vilchez1988} and the emission-line-ratio function \citep{Agostino2019}, but provides a simpler and more transparent continuous measure of departure from star-formation excitation.

In the field (PHANGS--MUSE), spaxels cluster tightly around the star-forming sequence, with roughly half lying below the Ka03 boundary. By contrast, cluster galaxies in MAUVE--MUSE contain only $\sim$20\% star-forming spaxels in the [\NII]-based diagram and exhibit a pronounced tail extending to $d_{\mathrm{SF}} \sim +0.5$--$+1.0$~dex, consistent with excitation from evolved stars, \HII\ leakage, or shocks. The [\SII]-based diagram shows a similarly broad distribution, with MAUVE--MUSE galaxies displaying a denser population of composite and LINER-like spaxels than the field sample. Together, these trends confirm that cluster environments drive systematic departures from pure star-forming excitation, preferentially enhancing low-ionization emission.

% Importantly, strong AGN are not present in either sample, suggesting that these harder ionization signatures arise not from nuclear activity but from extended, non-nuclear processes in the disk. 
Spatially correlated enhancements in [\NII]/\Ha, [\SII]/\Ha, and [\OIII]/\Hb\ are observed in the extraplanar features of NGC~4522, possibly due to environment-driven mechanisms such as shocks or turbulent mixing layers. However, these elevated line ratios are also evident in the aggregate cluster population, implying that such processes may be widespread. To determine the dominant secular and/or environment-related drivers of these excitation differences across the sample, it is essential to examine the gas kinematics in different ionization regimes and environmental contexts.

\subsection{Ionization Conditions and Kinematics}\label{sec:ha_sigma}

\begin{figure*}
    \centering
    \includegraphics{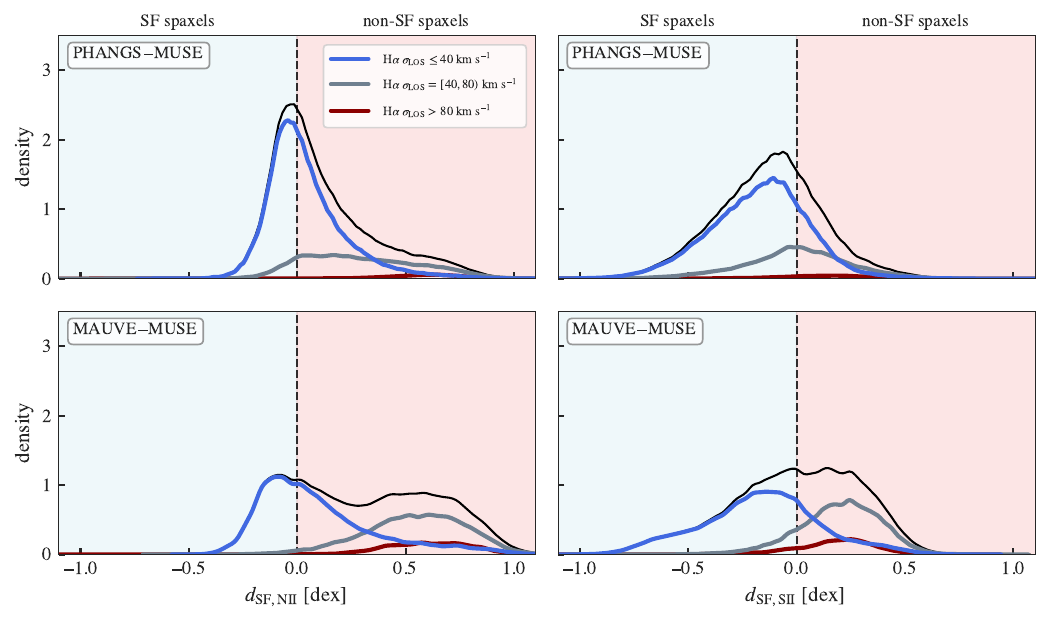}
    \caption{
    Distributions of orthogonal distance from the star-formation boundaries, $d_{\rm SF, NII}$ (left; using Ka03 in [\NII] BPT) and $d_{\rm SF, SII}$ (right; Ke01 [\SII] BPT), for the field (PHANGS--MUSE; top row) and cluster (MAUVE--MUSE; bottom row) samples. Spaxels are subdivided by \Ha~ velocity dispersion into three bins ($\sigma_{\rm LOS} \leq 40$kms$^{-1}$, $40 \leq \sigma_{\rm LOS} < 80$kms$^{-1}$, and $\sigma_{\rm LOS} \geq 80$kms$^{-1}$). Black curves show kernel density estimates for the full spaxel distributions. Spaxels are weighted such that each galaxy contributes equally. Positive values denote ratios offset from the star-forming sequence toward harder ionization regimes.
    }
    \label{fig:ionization_distance_kde}
\end{figure*}
In this section, we examine the relationship between \Ha~ velocity dispersion along the line-of-sight (\Ha~$\sigma_{\rm LOS}$) and gas ionization conditions to further distinguish the increased DIG prominence due to the relative decline in photoionization within quenching disks versus enhanced shock excitation due to ICM-ISM interaction.

Figure~\ref{fig:ionization_distance_kde} presents the distributions of orthogonal distance from the star-formation boundaries, $d_{\rm SF, NII}$ and $d_{\rm SF, SII}$, for spaxels in the PHANGS--MUSE and MAUVE--MUSE samples. These are shown separately for three H$\alpha$ velocity dispersion bins: $\sigma_{\rm LOS} \leq 40$kms$^{-1}$, $40 \leq \sigma_{\rm LOS} < 80$kms$^{-1}$, and $\sigma_{\rm LOS} \geq 80$kms$^{-1}$.

In the PHANGS--MUSE sample (top panels), both $d_{\rm SF, NII}$ and $d_{\rm SF, SII}$ peak near zero, with most spaxels lying below or close to the star-forming boundaries across all but the highest dispersions. This indicates that field disks are dominated by photoionization, with only a minority contribution from harder sources.

In contrast, the MAUVE--MUSE cluster galaxies (bottom panels) show a systematic positive shift and a larger overall fraction of high-dispersion spaxels. Spaxels with $\sigma_{\rm LOS} > 60$kms$^{-1}$ are displaced to higher $d_{\rm SF}$ values, peaking around 0.2–0.3 dex for [\SII]/\Ha and extending up to $\sim0.6$ dex for [\NII]/\Ha. This trend indicates that dynamically disturbed gas in cluster environments departs significantly from pure star formation excitation.

To understand the observed correlation between high line-of-sight velocity dispersions and harder ionization conditions, we now directly compare the distribution of  \Ha~$\sigma_{\rm LOS}$ in field and cluster galaxies as function of ionization state. If external processes such as ram pressure stripping are actively driving turbulence and shocks, we expect a significant excess of high-\Ha~$\sigma_{\rm LOS}$ spaxels in the cluster sample, particularly among regions classified as non-star-forming.

\begin{figure*}
    \centering
    \includegraphics{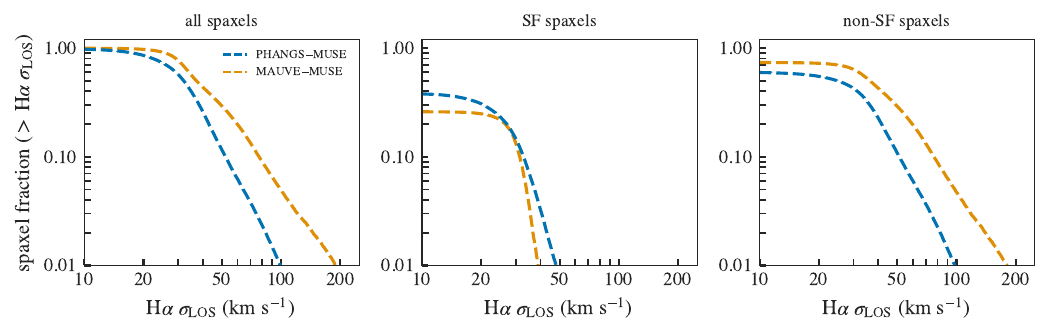}
    \caption{
    CCDFs of \Ha~line-of-sight velocity dispersion (\Ha~$\sigma_{\rm LOS}$) for the field (PHANGS--MUSE) and cluster (MAUVE--MUSE) galaxy samples. Left: all spaxels; middle and right panels: spaxels classified as star-forming and non-star-forming using the Ka03 boundary in the [\NII] BPT, respectively. Spaxels are weighted such that each galaxy contributes equally to the distribution.}
    \label{fig:ha_sigma_ecdf}
\end{figure*}

Figure~\ref{fig:ha_sigma_ecdf} shows the complementary cumulative distribution functions (CCDFs) of H$\alpha$ velocity dispersion for the PHANGS--MUSE and MAUVE--MUSE samples, split into all spaxels (left), star-forming spaxels ($d_{\rm SF, NII}<0$; middle), and non–star-forming spaxels ($d_{\rm SF, NII}\geq0$; right). We adopt the [\NII]-based BPT diagnostic since it highlights the strongest field–cluster contrasts, although this choice does not affect our conclusions.

The overall balance of ionization sources is markedly different. PHANGS–MUSE galaxies contain $\sim40$\% star-forming and $\sim60$\% non–star-forming spaxels, whereas in MAUVE–MUSE the split is $\sim25$\% versus $\sim75$\%. This shift strongly shapes the velocity dispersion distributions.

For all spaxels (left), cluster galaxies show a much higher incidence of dynamically disturbed gas. At $\sigma_{\rm LOS} > 40$ km s$^{-1}$, 44\% of MAUVE–MUSE spaxels exceed this threshold compared to only 26\% in PHANGS–MUSE. At $\sigma_{\rm LOS} > 100$ km s$^{-1}$, 5\% of spaxels are above this level in cluster galaxies versus 1\% in the field.

Crucially, the $\sigma_{\rm LOS}$ distributions for star-forming spaxels are very similar in field and cluster environments. The median H$\alpha$ $\sigma_{\mathrm{LOS}}$ for star-forming regions is $27~\mathrm{km~s^{-1}}$ in PHANGS--MUSE and $30~\mathrm{km~s^{-1}}$ in MAUVE--MUSE, with narrow interquartile ranges ($12~\mathrm{km~s^{-1}}$ and $6~\mathrm{km~s^{-1}}$, respectively) and 90\% ranges ($\sim31~\mathrm{km~s^{-1}}$ and $\sim18~\mathrm{km~s^{-1}}$). These similar, narrow distributions demonstrate that turbulence within H\textsc{ii} regions is not systematically enhanced in clusters.

The excess of high-dispersion gas in MAUVE–MUSE arises almost entirely from non–star-forming spaxels, which exhibit substantially broader kinematics (right panel). The median $\sigma_{\rm LOS}$ increases from $36~\mathrm{km~s^{-1}}$ in the field to $44~\mathrm{km~s^{-1}}$ in clusters. More notably, the 50\% range (IQR) widens from $17~\mathrm{km~s^{-1}}$ to $28~\mathrm{km~s^{-1}}$, and the 90\% range expands from $58~\mathrm{km~s^{-1}}$ in the field to $85~\mathrm{km~s^{-1}}$ in the cluster environment. This behavior is consistent with the increased prominence of diffuse ionized gas once star formation is suppressed.

\section{Discussion}\label{sec:discussion}

Our results provide clear evidence that the ISM in MAUVE--MUSE galaxies exhibits harder ionization conditions and systematically elevated \Ha~line-of-sight velocity dispersions relative to PHANGS--MUSE galaxies. The spatially resolved BPT diagrams show that cluster galaxies extend $\sim0.5{-}1.0$ dex above the star-forming locus (Figure~\ref{fig:sf_distance_bpt}), such that $\sim75$\% of spaxels fall outside the star-forming regime ($d_{\rm SF,NII}\geq0$), compared to $\sim60$\% in the field.

This result is driven by the quantitative differences in the line ratio distributions (Table~\ref{tab:line_ratio_stats}, Figure~\ref{fig:line_ratios_phangs_mauve}). For [\NII]/\Ha, the median increases from 0.50 in PHANGS–MUSE to 0.75 in MAUVE–MUSE, while the interquartile range (IQR = P$_{75}$–P$_{25}$) broadens from 0.30 to 0.76 and the 90\% range (P$_{95}$–P$_5$) from 1.07 to 1.70. Similar shifts are seen in [\SII]/\Ha\ (median 0.49 $\rightarrow$ 0.57; IQR 0.27 $\rightarrow$ 0.35; 90\% range 0.66 $\rightarrow$ 0.80) and [\OIII]/\Hb\ (median 0.68 $\rightarrow$ 1.04; IQR 0.64 $\rightarrow$ 1.42; 90\% range 2.16 $\rightarrow$ 3.56).

These ionization differences are closely linked to broadened ionized gas kinematics in cluster galaxies. In PHANGS--MUSE, the $d_{\rm SF,NII}$ and $d_{\rm SF,SII}$ distributions remain peaked below zero across all velocity bins, showing that even intermediate-dispersion gas is dominated by photoionization. MAUVE--MUSE galaxies, by contrast, exhibit a marked positive shift for spaxels with $\sigma_{\rm LOS}>40$ km s$^{-1}$, with peaks around $d_{\rm SF}\sim+0.5$ dex and extended wings reaching $\sim+1$ dex (Figure \ref{fig:ionization_distance_kde}). This change is mirrored in the CCDFs (Figure~\ref{fig:ha_sigma_ecdf}), where nearly half of all cluster spaxels lie above $\sigma_{\rm LOS}=40$ km s$^{-1}$ compared to only a quarter of spaxels in the field. At higher dispersions the gap between the cluster and field samples widens further, with most of the excess in cluster galaxies arising from non–star-forming spaxels in the $40{-}80$ km s$^{-1}$ range—consistent with enhanced visibility of DIG once star formation is suppressed.

The star-forming component itself shows little difference between environments. Only 3\% of PHANGS–MUSE star-forming spaxels and 1\% of MAUVE–MUSE star-forming spaxels exceed 40 km s$^{-1}$, and both samples fall to 0\% by 80 km s$^{-1}$. This indicates that the turbulence of ionized gas is comparable in both environments. The observed cluster–field contrast therefore arises largely from the relative balance of star-forming and non–star-forming gas, rather than from changes in the turbulence of star-forming regions themselves.

However, MAUVE--MUSE galaxies also contain a significant sub-dominant population of high-dispersion spaxels absent in PHANGS--MUSE. We find 5\% of all cluster spaxels exceed $\sigma_{\rm LOS}=100$ km s$^{-1}$, compared to 1\% in the field. These extreme values occur almost exclusively in non–star-forming regions with elevated line ratios, consistent with a contribution from environment-driven shocks or turbulent mixing layers.

Taken together, this points to two regimes: (1) DIG-like emission dominating at moderate dispersions (\Ha~$\sigma_\mathrm{LOS} = 40 - 100~\mathrm{km~s^{-1}}$), driving the main cluster–field difference, and (2) a smaller but non-negligble contribution from shock-ionized gas at the highest dispersions (\Ha~$\sigma_\mathrm{LOS} > 100~\mathrm{km~s^{-1}}$).

The MAUVE--MUSE sample falls below the main sequence of star formation (Figure \ref{fig:global_main_sequence}), a direct consequence of environmental processing in Virgo. We argue that the most likely scenario is that the observed difference in ionization conditions is primarily a result of this quenching process. In field galaxies, the luminosity-weighted emission is dominated by high surface brightness \HII~ regions. In cluster galaxies, as environmental mechanisms (e.g., ram pressure stripping, starvation) suppress star formation, this \HII~component fades. Consequently, the pervasive DIG, characterized by elevated line ratios and turbulence, becomes the dominant visible component of the ISM. Direct environmental excitation, such as shock heating from the ICM wind, appears to be a secondary effect, manifesting primarily in the high-velocity dispersion tail (\Ha~$\sigma_\mathrm{LOS} > 100~\mathrm{km~s^{-1}}$).

% Taken together, the data support a picture in which environmental transformation of the ISM is driven primarily by the suppression of star formation, which unveils a widespread DIG component with moderate dispersions ($40{-}80$ km s$^{-1}$). A secondary contribution from shocks accounts for the excess of very high dispersion ($\gtrsim100$ km s$^{-1}$) spaxels, particularly in galaxies experiencing the most extreme ram pressure conditions.

This interpretation is consistent with our case studies. In NGC~4501, an early-stage infaller, hard ionization and moderate velocity dispersions are found throughout the disk as star formation fades, consistent with DIG dominance. In contrast, NGC~4522, being close to peak ram pressure, exhibits extended shock-ionized tails with compact star-forming knots embedded in the stripped gas (Figures \ref{fig:NGC4501_emline_ratios} and \ref{fig:NGC4522_emline_ratios}). This evolution from DIG emergence during early quenching to shock-dominated excitation during peak stripping matches theoretical expectations \citep{Tonnesen2012, Boselli2022} and likely unfolds on $\lesssim~\mathrm{few}~100$ Myr timescales as galaxies traverse the cluster. Fully understanding this picture and the timescales involved will be the focus of future work.

% We see examples of increased DIG visibility and environment-driven shocks respectively in our two case studies:  NGC~4501 (an early-stage infaller, Figures \ref{fig:NGC4501_emline_ratios} and \ref{fig:photoion_maps}) and NGC~4522 (a late-stage infaller, Figures \ref{fig:NGC4522_emline_ratios} and \ref{fig:photoion_maps}). This evolution from initial, confined gas perturbations with enhanced line ratios as photoionization fades, to extended shock-ionized tails during the peak stripping phase is in line with theoretical expectations \citep{Tonnesen2012, Boselli2022}. It underscores that environmental ionization signatures are strongest during active stripping and may evolve rapidly on $\sim100$~Myr timescales as galaxies traverse the cluster. 

\section{Conclusions}\label{sec:conclusions}
We have presented high-resolution IFS data from the MAUVE--MUSE survey of Virgo Cluster galaxies to investigate how dense environments impact ionized gas properties. Our main conclusions are:

\begin{enumerate}
        \item Ionization structure is spatially correlated with environmental influence in our ram pressure stripping case studies: NGC~4501 (early infall, DIG emergence, Figure \ref{fig:NGC4501_emline_ratios}) and NGC~4522 (close to peak ram pressure, shock-ionized tails with embedded star-forming knots, Figure \ref{fig:NGC4522_emline_ratios}).
        \item MAUVE–MUSE galaxies exhibit systematically elevated emission line ratios compared to field disks, with higher medians of [\NII]/\Ha\ (0.75 vs.\ 0.50), [\SII]/\Ha\ (0.57 vs.\ 0.49), and [\OIII]/\Hb\ (1.04 vs.\ 0.68), and broader percentile ranges indicating a stronger contribution from non-\HII\ region ionization processes (Table~\ref{tab:line_ratio_stats}; Figure~\ref{fig:line_ratios_phangs_mauve}). These shifts primarily reflect the increased prominence of DIG as star formation is suppressed.
        \item Spatially resolved BPT diagrams show that 74\% of spaxels in MAUVE--MUSE galaxies (cluster) lie above the star-forming boundary ($d_{\rm SF,NII}\geq0$), compared to 61\% in PHANGS--MUSE (field), with offsets in the MAUVE--MUSE galaxies of $\sim0.5{-}1.0$ dex (Figure~\ref{fig:sf_distance_bpt}).
        \item Elevated velocity dispersions in cluster galaxies are primarily associated with non–star-forming spaxels. 44\% of all cluster spaxels exceed $\sigma_{\rm LOS}=40$ km s$^{-1}$ (vs.\ 26\% in the field), in both cases this is dominated by non-star forming gas while the star-forming component shows little kinematic difference (Figures \ref{fig:ionization_distance_kde} and ~\ref{fig:ha_sigma_ecdf}).
        \item The dominant environmental effect is the suppression of star formation, which shifts the balance of ionized gas toward DIG. This is most evident in the intermediate velocity regime: in MAUVE--MUSE, 35\% of all spaxels lie in the range $40{-}80$ km s$^{-1}$ (vs.\ 24\% in PHANGS--MUSE), consistent with enhanced DIG dominating once compact \HII\ regions decline in prominence (Figure \ref{fig:ha_sigma_ecdf}).
        \item At the highest dispersions, only cluster galaxies host a significant tail: 5\% of MAUVE--MUSE spaxels exceed $\sigma_{\rm LOS}>100$ km s$^{-1}$ compared to just $\sim1$\% in the field. These high dispersion values are confined to non–star-forming spaxels and are consistent with a subdominant but important contribution from shocks or turbulent mixing layers driven by ICM interaction (Figure \ref{fig:ha_sigma_ecdf}).
\end{enumerate}

These results suggest that the primary impact of the cluster environment on the ionized ISM is the suppression of star formation, which reveals DIG as the dominant ionized phase across galaxy disks, while shock-driven excitation associated with active stripping contributes a secondary, non-negligible high-dispersion component.

\section{Data Availability}
The full set of MAUVE--MUSE emission line flux maps will be made available as part of main public data release once the survey is completed (exp. 2027). Please contact the corresponding author if you desire data access prior to this release.

\section{Acknowledgments}
The majority of this work was conducted on the traditional territory of the T'Sou-ke and Lekwungen peoples. The authors acknowledge and respect the T'Sou-ke, Songhees, Esquimalt and WS\'{A}NE\'{C} Nations whose historical relationships with the land continue to this day.

This work is carried out as part of the MAUVE collaboration (https://mauve.icrar.org/) and is based on observations collected at the European Southern Observatory (ESO) under ESO programmes 110.244E. This paper makes use of services that have been provided by AAO Data Central  (datacentral.org.au). The authors acknowledge the use of the Canadian Advanced Network for Astronomy Research (CANFAR) Science Platform operated by the Canadian Astronomy Data Center (CADC) and the Digital Research Alliance of Canada (DRAC), with support from the National Research Council of Canada (NRC), the Canadian Space Agency (CSA), CANARIE, and the Canadian Foundation for Innovation (CFI).

% ADD ACKNOWLEDGEMENT HERE
L.C. acknowledges support from the Australian Research Council’s Discovery Project funding scheme (DP210100337).
AC acknowledges support by the National Research Foundation of Korea (NRF), No. RS-2022-NR070872 and 934  RS-2022-NR069020.
AR acknowledges the support by the Australian Research Council Centre of Excellence in Optical Microcombs for Breakthrough Science (project number CE230100006), with funding from the Australian Government. 
C.D.W. and KS acknowledge support from the Natural Sciences and Engineering Research Council of Canada (NSERC). C.D.W. acknowledges support from the Canada Research Chairs program (CRC-2022-00184). 
TK gratefully acknowledges the financial support from the Leverhulme Trust.
NZ is supported through the South African Research Chairs Initiative of the Department of Science and Technology and National Research Foundation.

% Here we acknowledge the key software used in this work:
% \label{sec:acknowledgments}
% \end{acknowledgments}


% \vspace{5mm}
\facilities{The Very Large Telescope (VLT); The Multi Unit Spectroscopic Explorer (MUSE)}


\software{
\href{http://www.astropy.org}{\texttt Astropy} \citep{astropy:2013, astropy:2018,larry_bradley_2020_4044744}, 
\href{https://github.com/sncosmo/extinction}{\texttt extinction}, 
\href{https://jiffyclub.github.io/palettable/}{\texttt palettable}, 
\href{https://matplotlib.org/}{\texttt Matplotlib} \citep{Hunter2007, 2020SciPy-NMeth}, 
\href{https://geckos-survey.github.io/gist-documentation/}{\texttt nGIST} \citep{Fraser-McKelvie2024}, 
\href{https://pandas.pydata.org/}{\texttt Pandas} \citep{mckinney-proc-scipy-2010}, 
\href{https://pypi.org/project/ppxf/}{\texttt pPXF} \citep{Cappellari2004,Cappellari2017,Cappellari2023}, 
\href{https://www.ascl.net/2507.015}{\texttt nGIST}, 
\href{https://scipy.org/}{SciPy} \citep{2020SciPy-NMeth}, 
\href{https://seaborn.pydata.org/index.html}{\texttt Seaborn} \citep{Waskom2021}, and 
\href{https://pypi.org/project/vorbin/}{\texttt Vorbin} \citep{Cappellari2003}}

\newpage
\appendix

\section{Emission Line Map Atlas}
\label{appendix:emlines_atlas}
Emission line data reduction is described in Section \ref{sec:Data}.

\begin{figure*}
    \centering
    \includegraphics{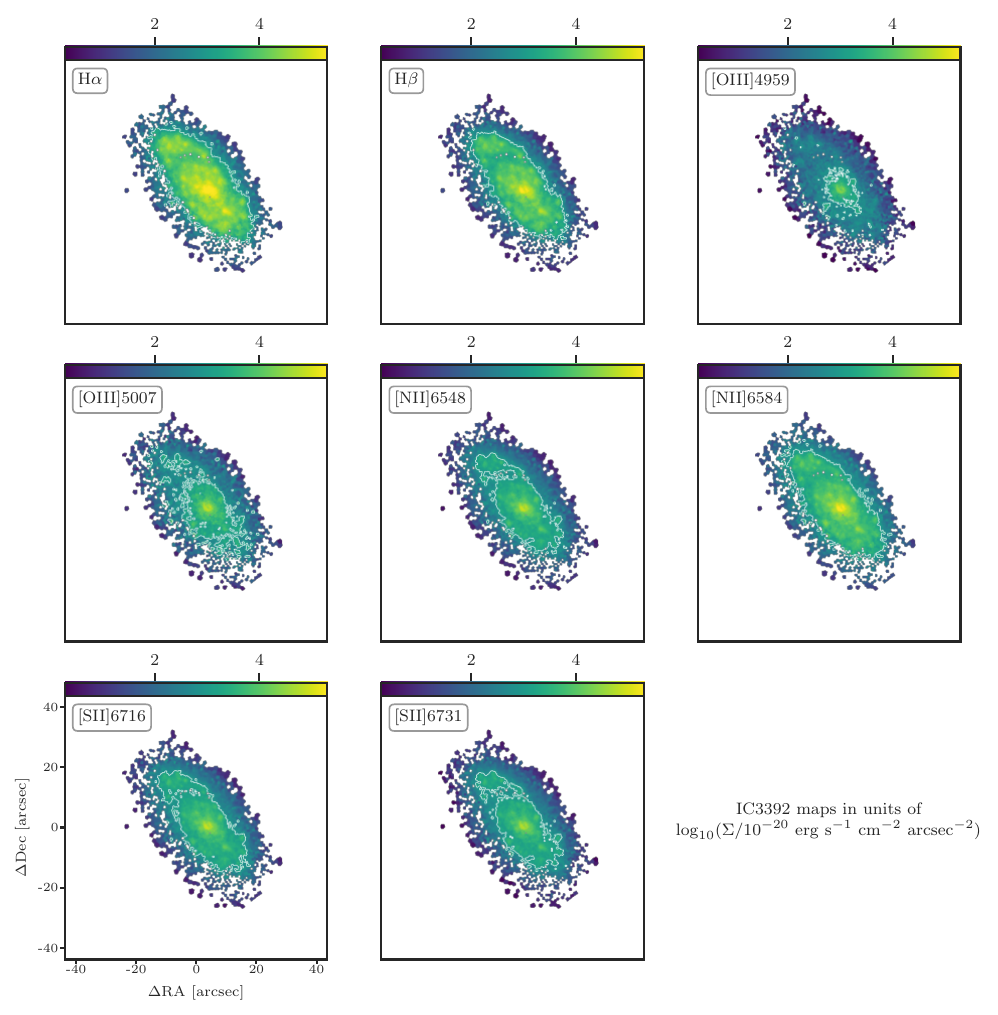}
    \caption{Same as Figure \ref{fig:emlines_ngc4501} but for IC~3392.}
\end{figure*}

\begin{figure*}
    \centering
    \includegraphics{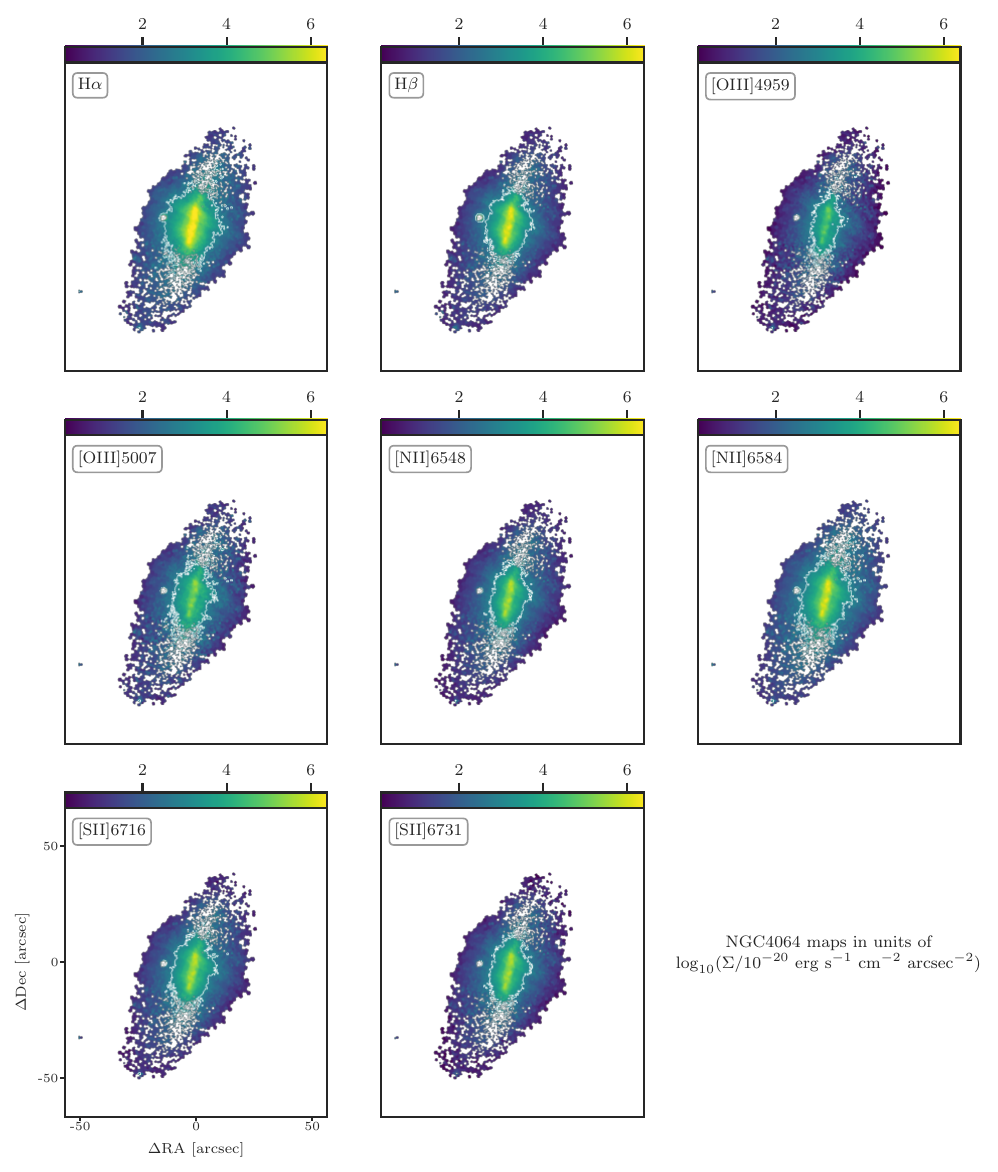}
    \caption{Same as Figure \ref{fig:emlines_ngc4501} but for NGC~4064.}
\end{figure*}

\begin{figure*}
    \centering
    \includegraphics[height=0.9\textheight]{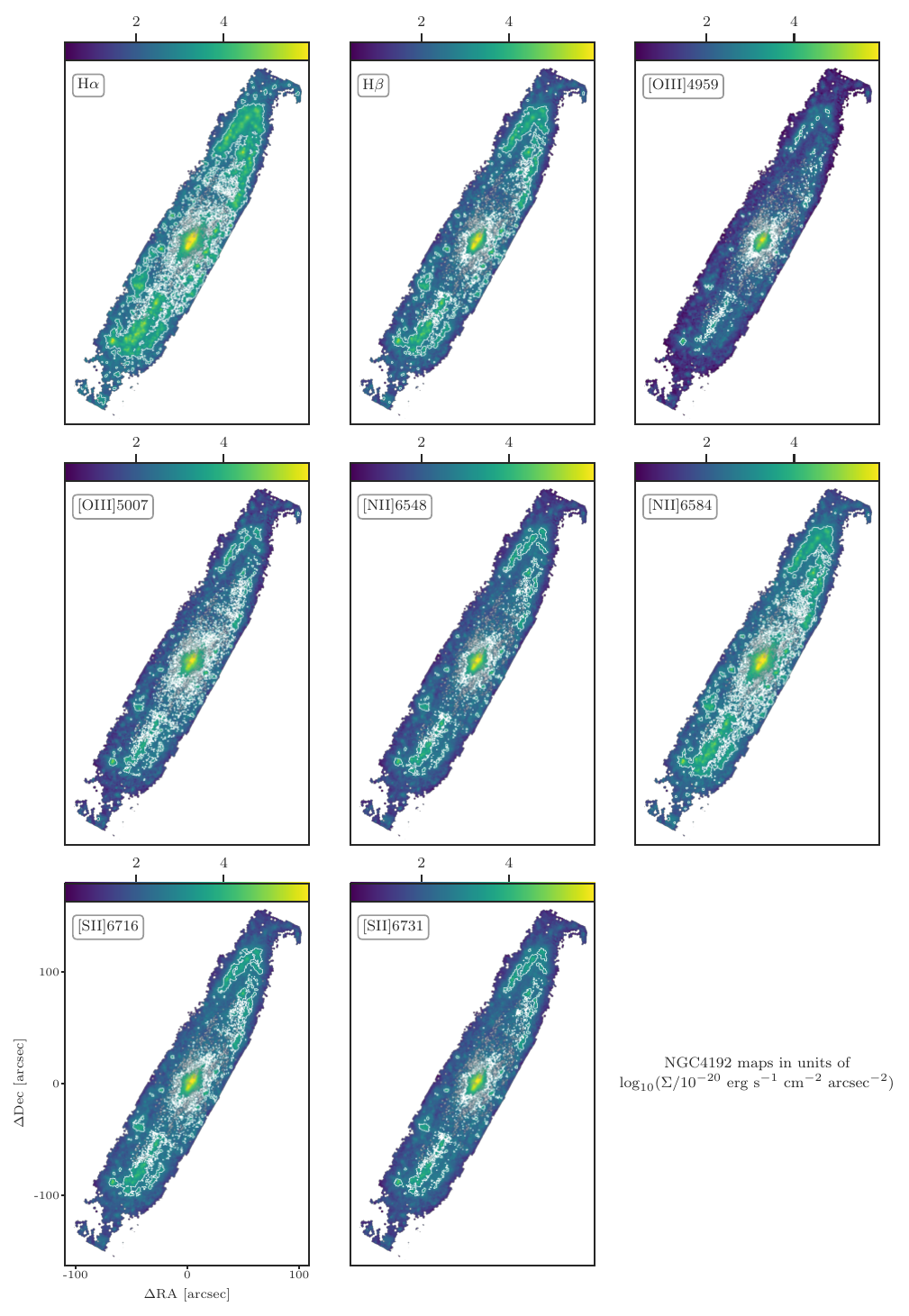}
    \caption{Same as Figure \ref{fig:emlines_ngc4501} but for NGC~4192.}
\end{figure*}

\begin{figure*}
    \centering
    \includegraphics{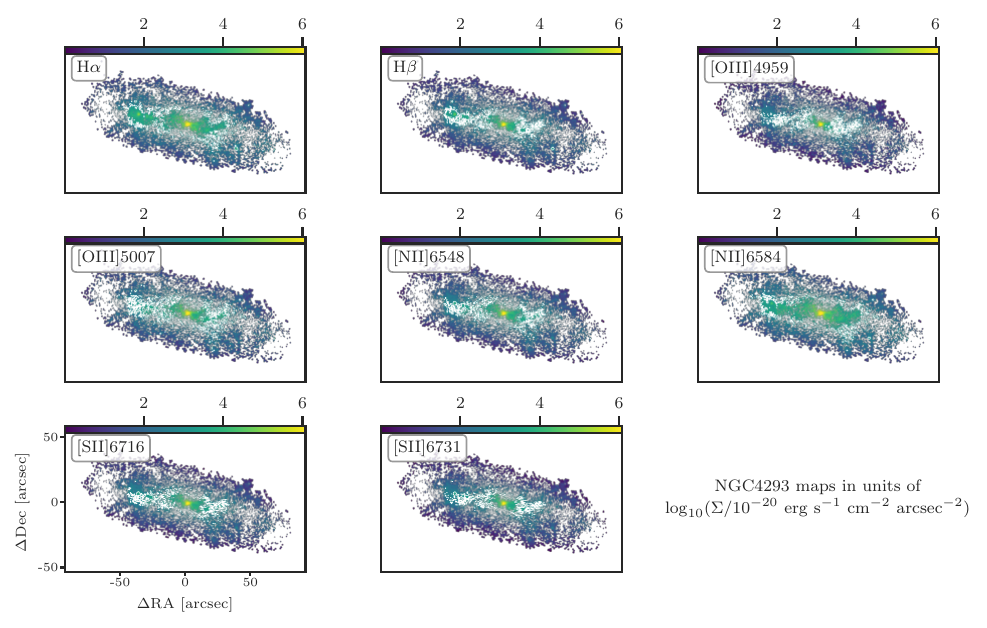}
    \caption{Same as Figure \ref{fig:emlines_ngc4501} but for NGC~4293.}
\end{figure*}

\begin{figure*}
    \centering
    \includegraphics{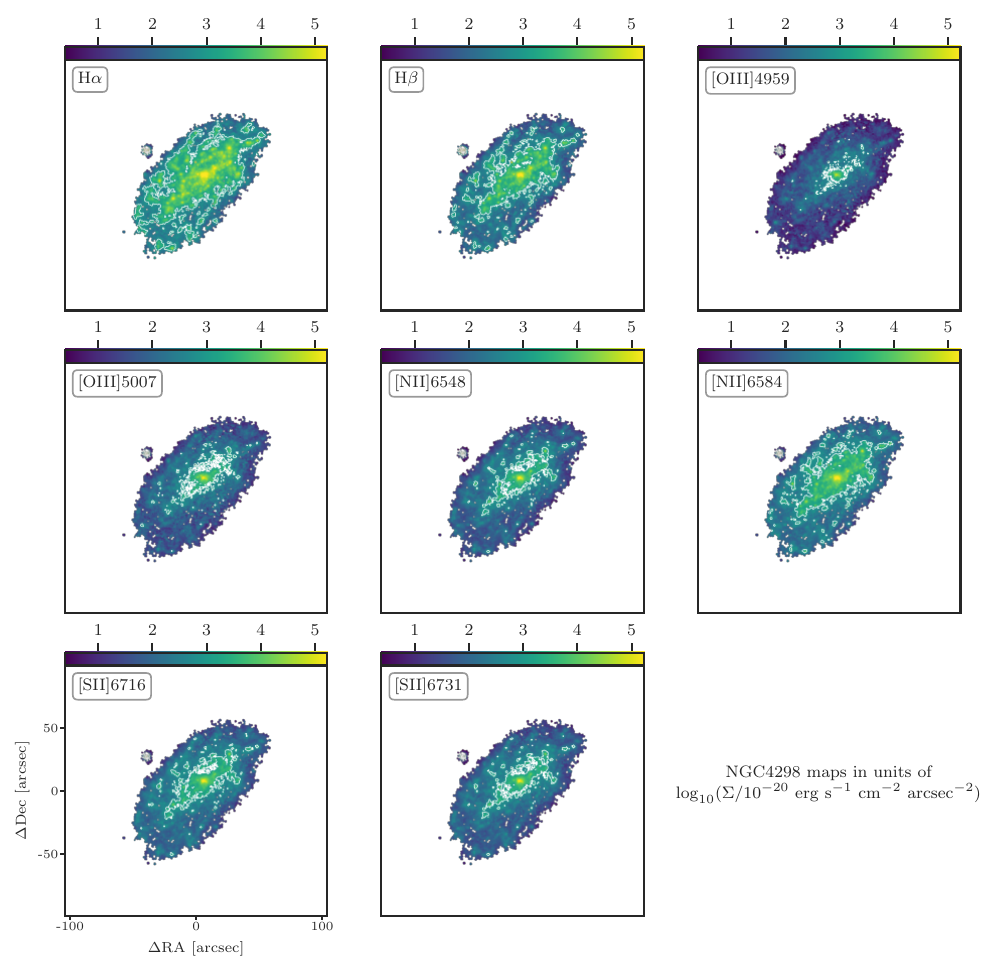}
    \caption{Same as Figure \ref{fig:emlines_ngc4501} but for NGC~4298.}
\end{figure*}

\begin{figure*}
    \centering
    \includegraphics{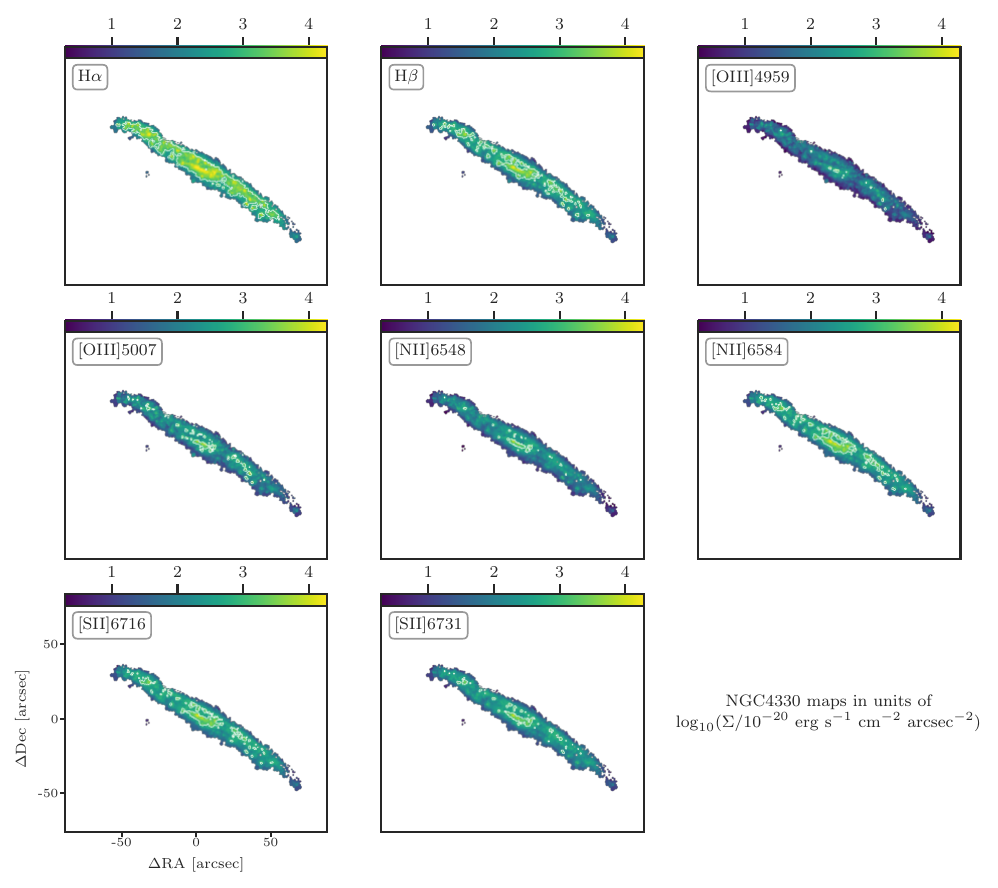}
    \caption{Same as Figure \ref{fig:emlines_ngc4501} but for NGC~4330.}
\end{figure*}

\begin{figure*}
    \centering
    \includegraphics{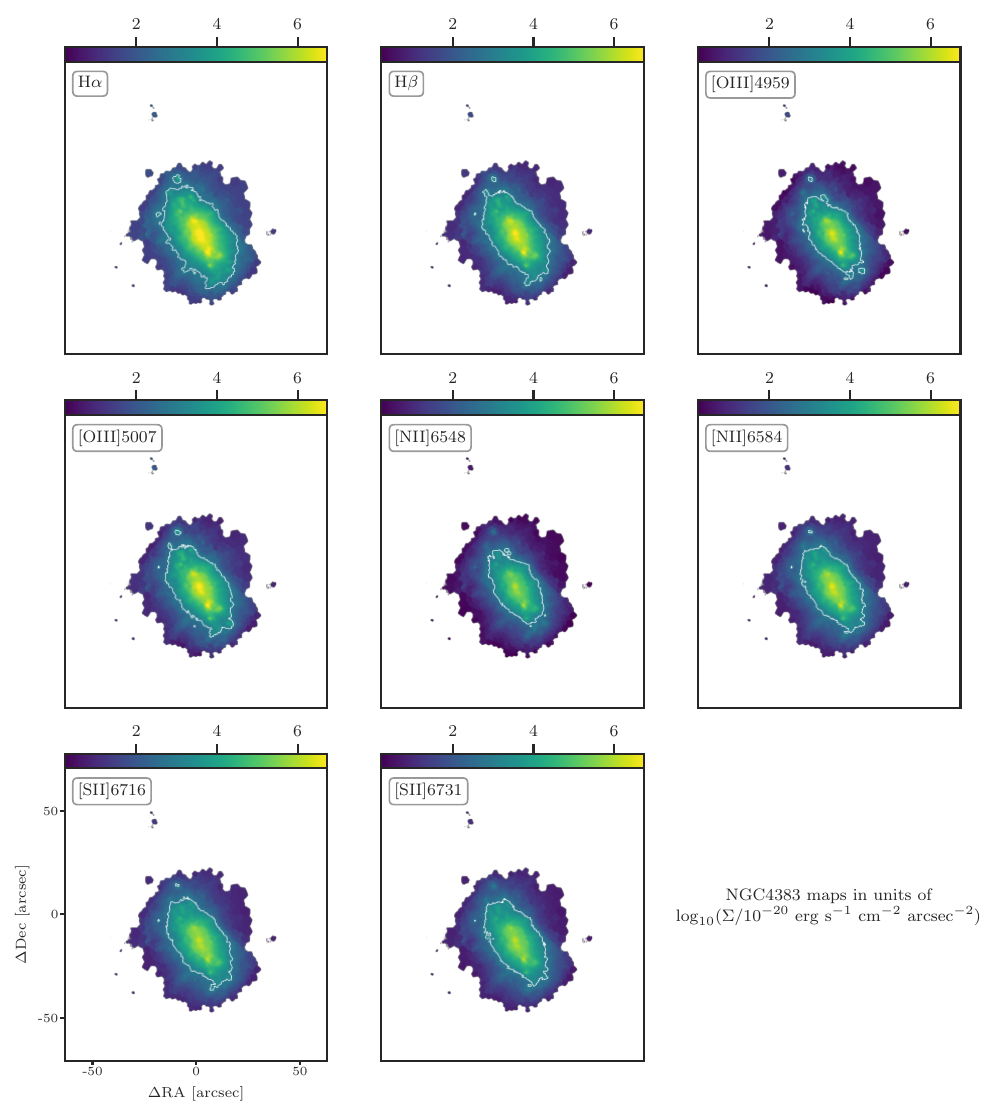}
    \caption{Same as Figure \ref{fig:emlines_ngc4501} but for NGC~4383.}
\end{figure*}

\begin{figure*}
    \centering
    \includegraphics{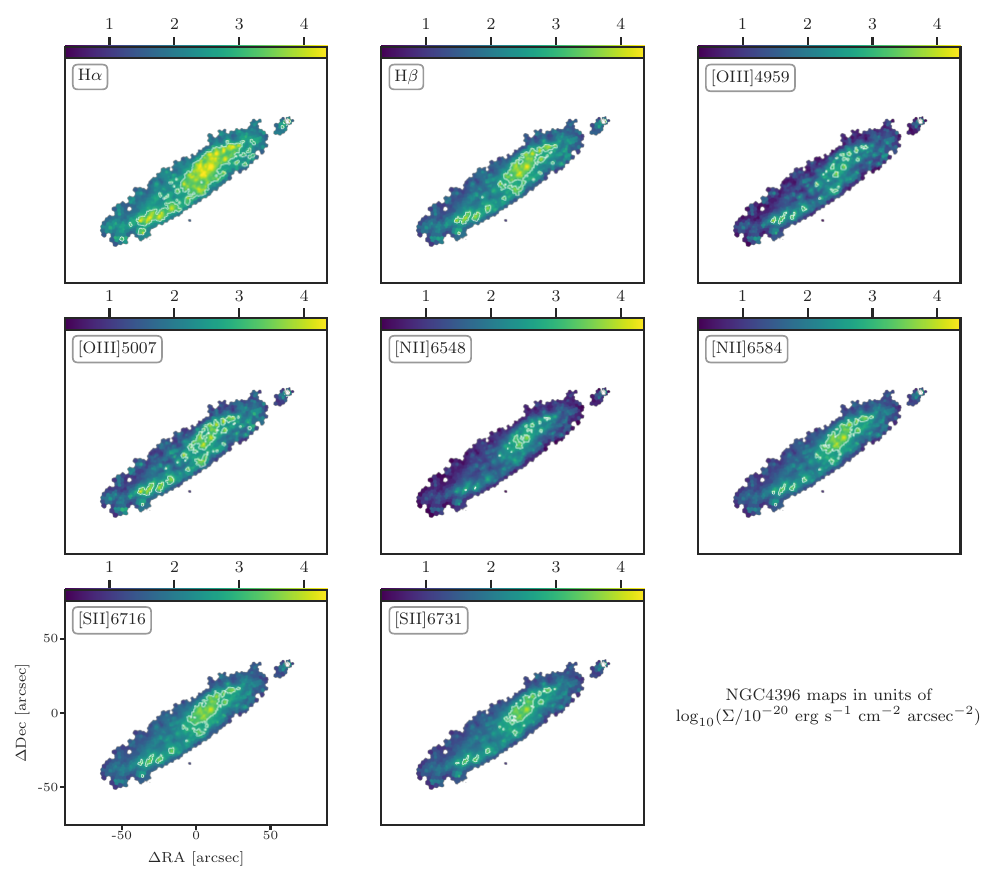}
    \caption{Same as Figure \ref{fig:emlines_ngc4501} but for NGC~4396.}
\end{figure*}

\begin{figure*}
    \centering
    \includegraphics{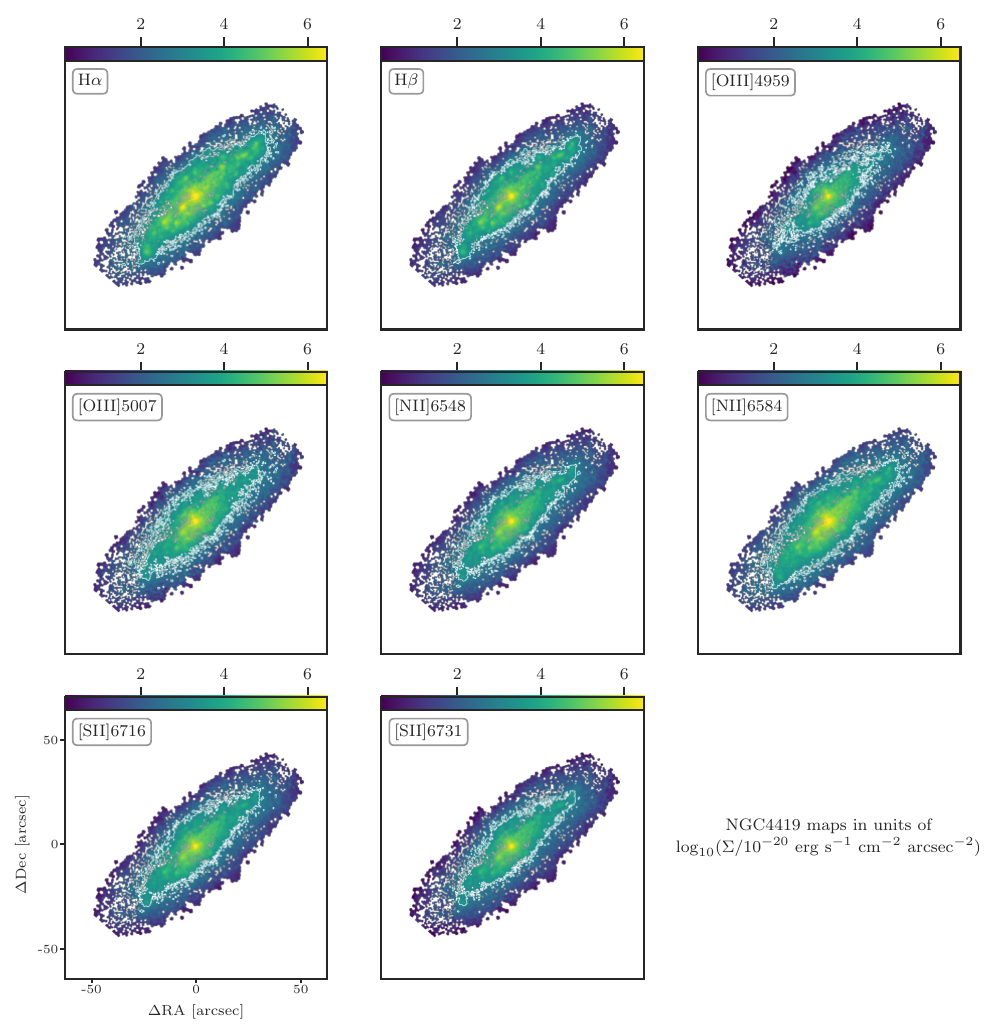}
    \caption{Same as Figure \ref{fig:emlines_ngc4501} but for NGC~4419.}
\end{figure*}

\begin{figure*}
    \centering
    \includegraphics{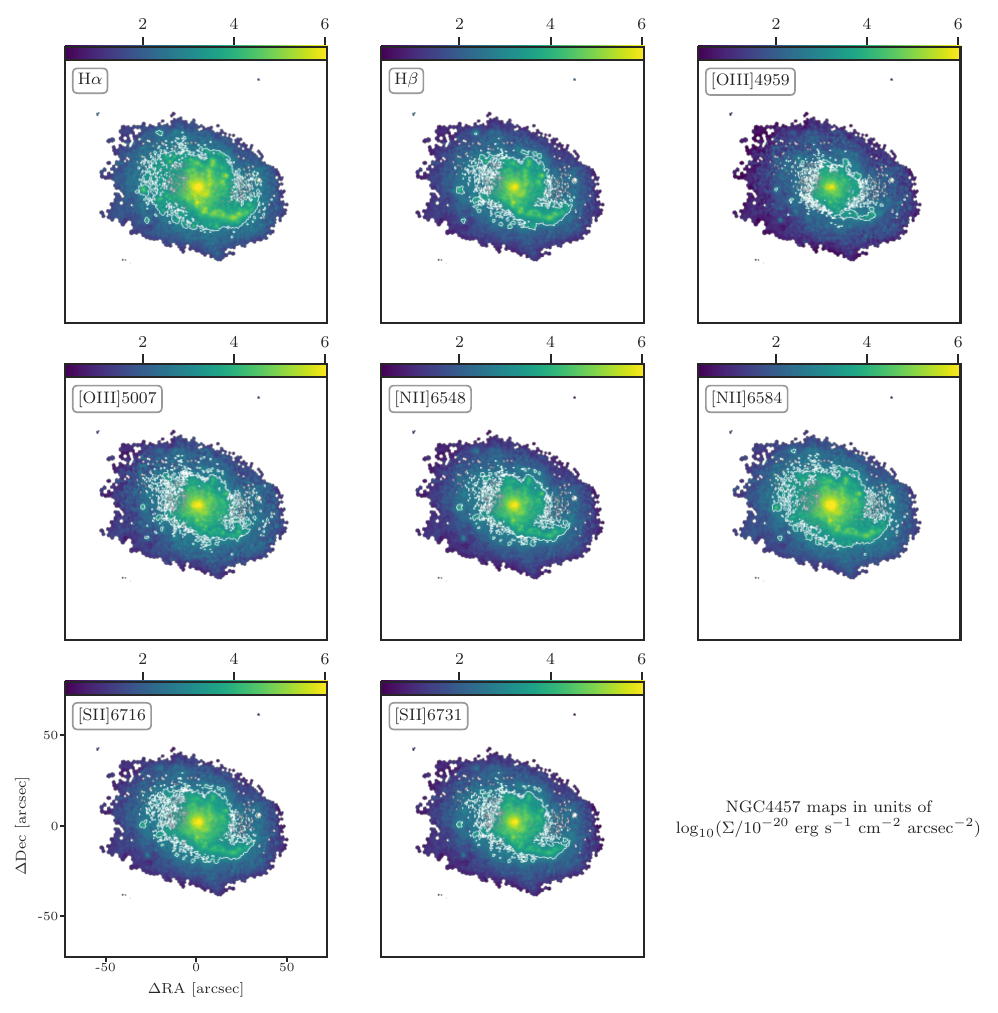}
    \caption{Same as Figure \ref{fig:emlines_ngc4501} but for NGC~4457.}
\end{figure*}

\begin{figure*}
    \centering
    \includegraphics{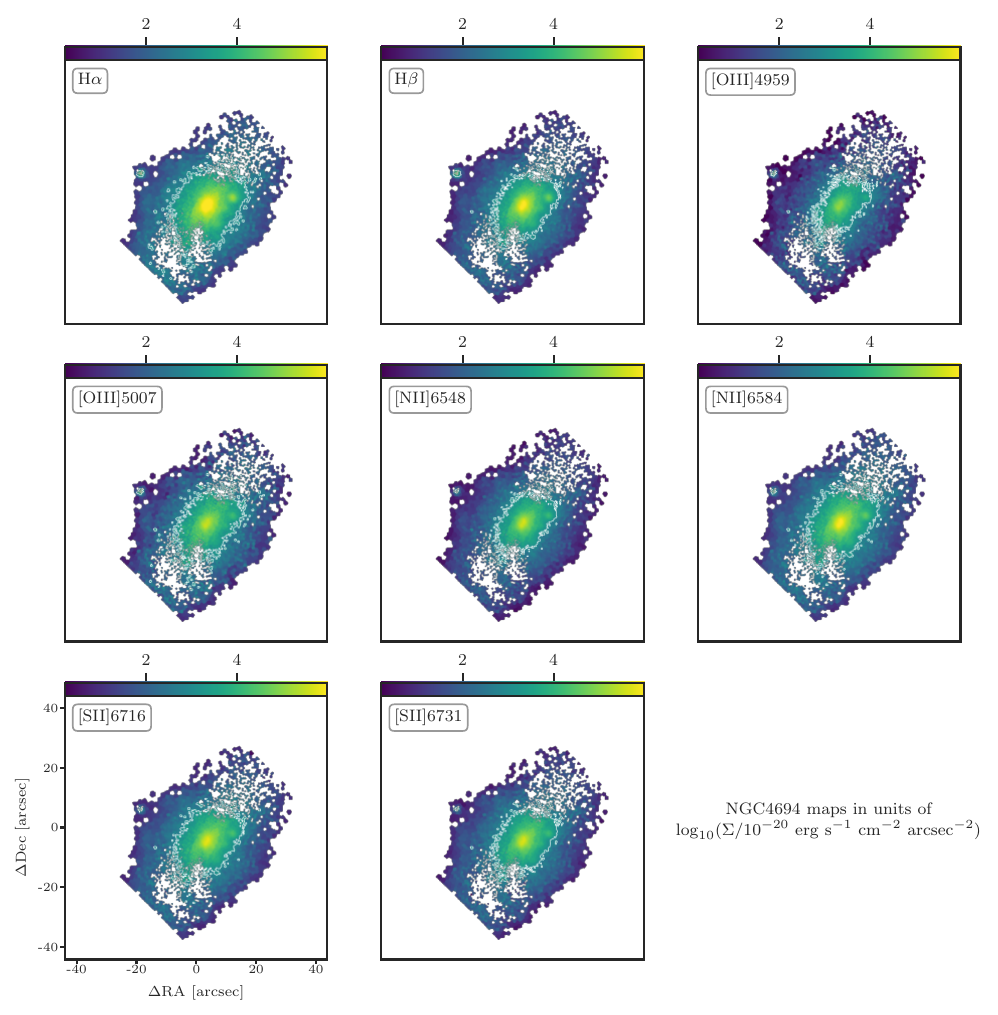}
    \caption{Same as Figure \ref{fig:emlines_ngc4501} but for NGC~4694.}
\end{figure*}

\begin{figure*}
    \centering
    \includegraphics[height=0.9\textheight]{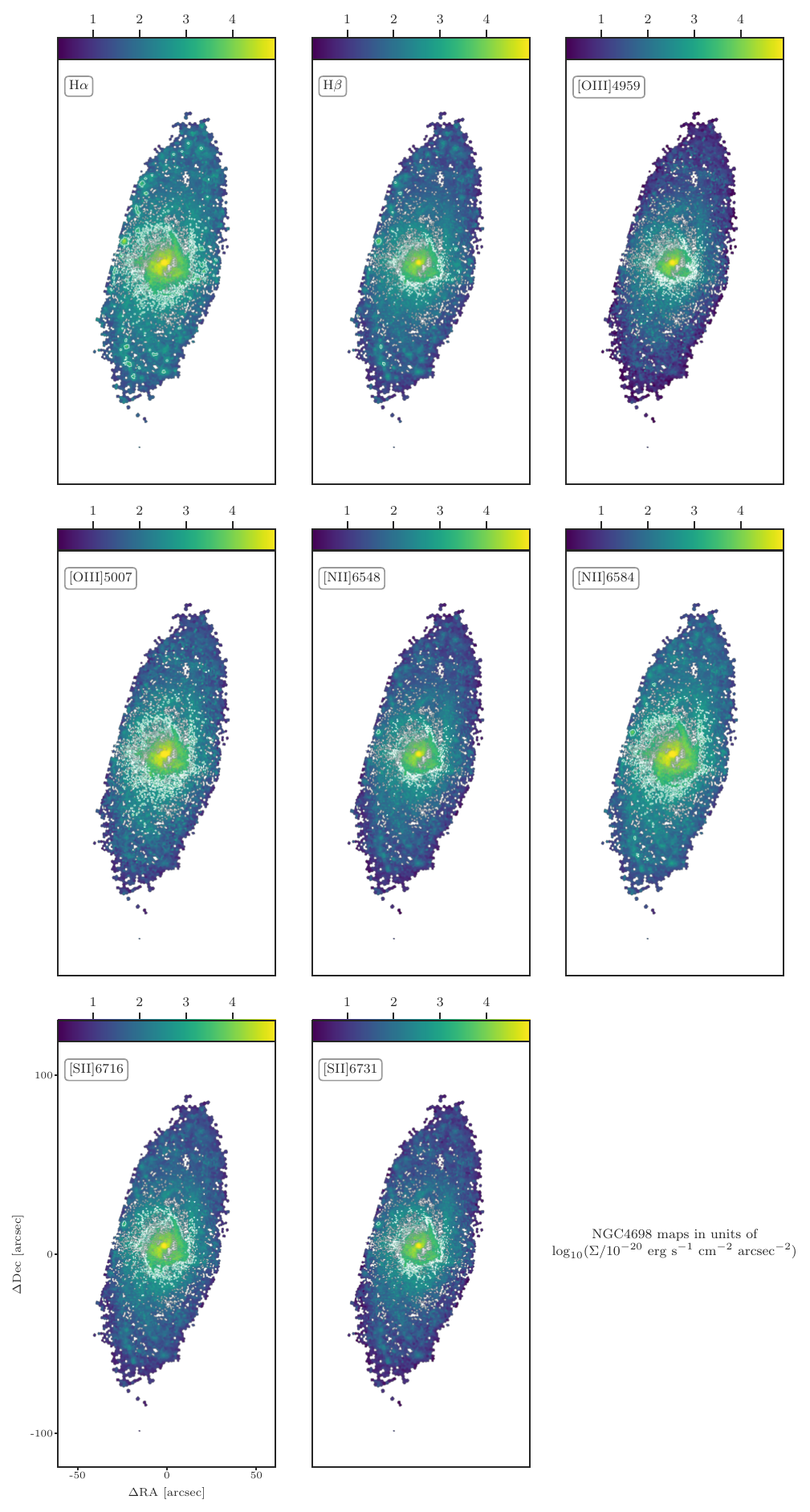}
    \caption{Same as Figure \ref{fig:emlines_ngc4501} but for NGC~4698.}
\end{figure*}

\section{Line Ratios Atlas}
\label{appendix:line_ratio_atlas}
See Section \ref{sec:emission_line_ratios} for details.

\begin{figure*}
    \centering
    \includegraphics{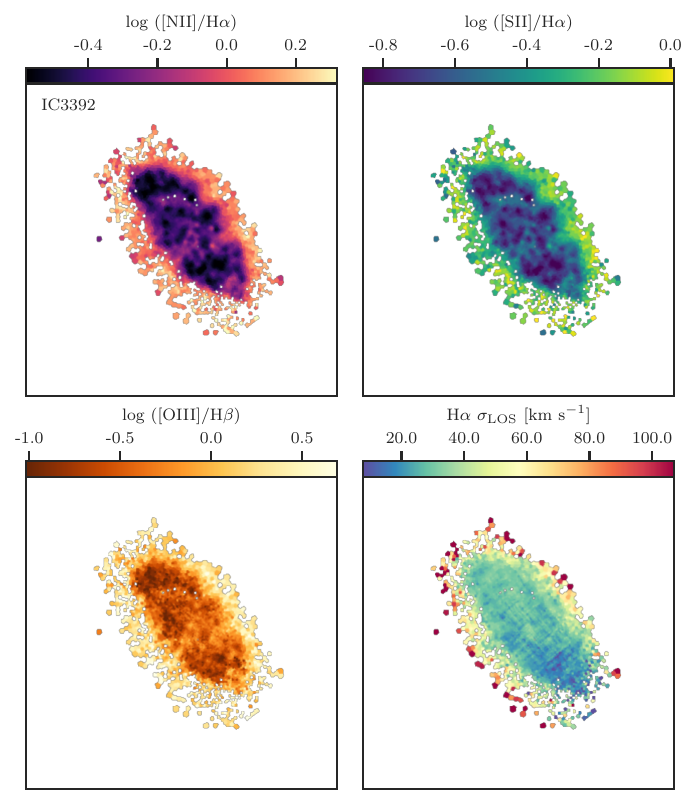}
    \caption{Same as Figure \ref{fig:NGC4501_emline_ratios} but for IC~3392.}
\end{figure*}

\begin{figure*}
    \centering
    \includegraphics[height=0.9\textheight]{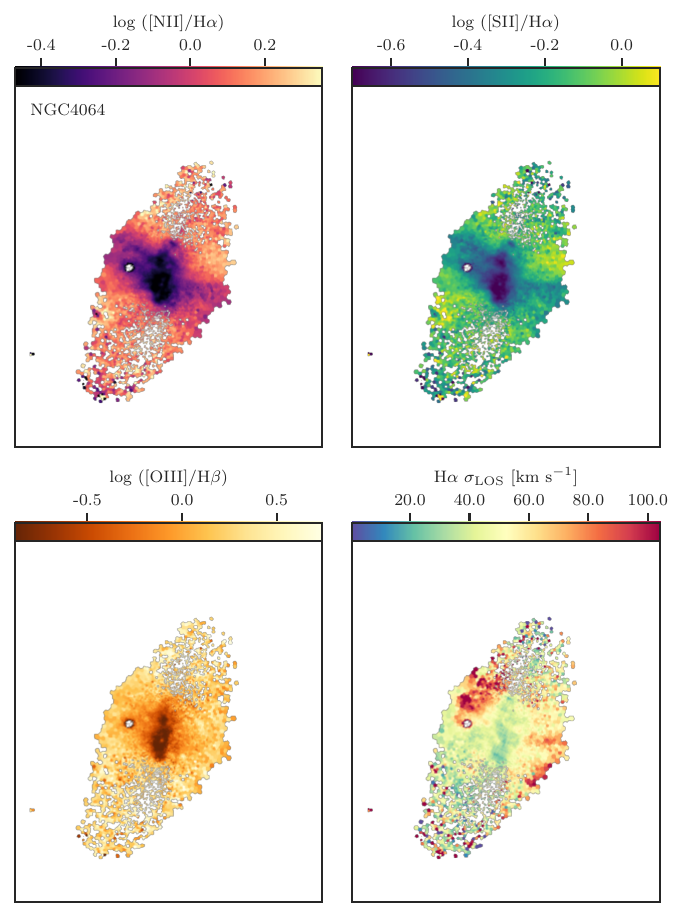}
    \caption{Same as Figure \ref{fig:NGC4501_emline_ratios} but for NGC~4064.}
\end{figure*}

\begin{figure*}
    \centering
    \includegraphics[height=0.9\textheight]{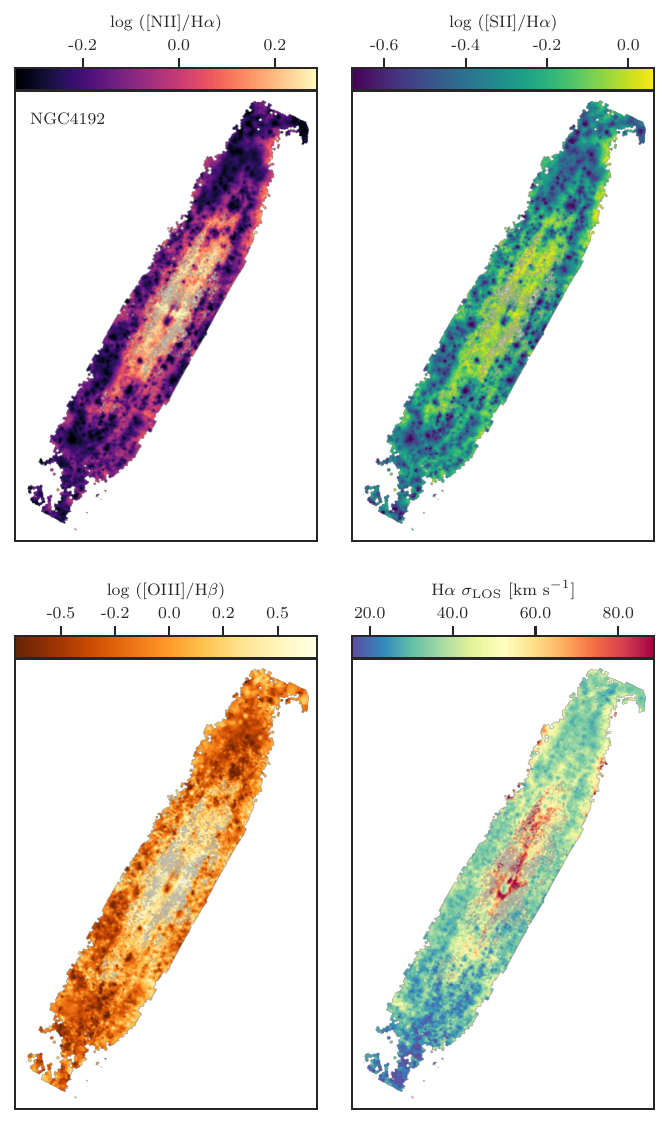}
    \caption{Same as Figure \ref{fig:NGC4501_emline_ratios} but for NGC~4192.}
\end{figure*}

\begin{figure*}
    \centering
    \includegraphics{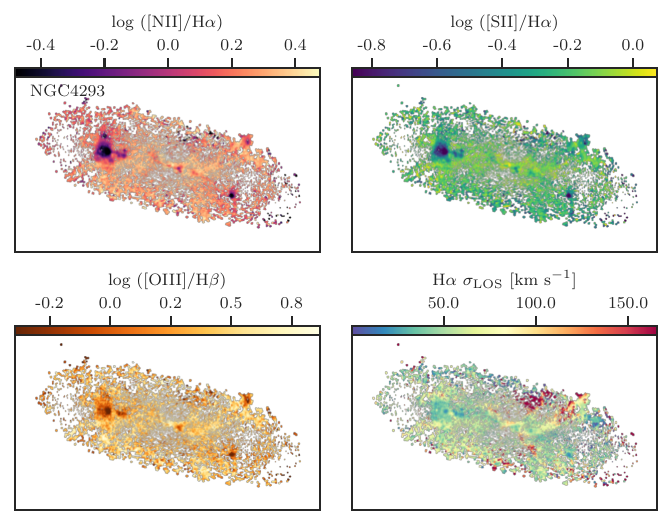}
    \caption{Same as Figure \ref{fig:NGC4501_emline_ratios} but for NGC~4293.}
\end{figure*}

\begin{figure*}
    \centering
    \includegraphics{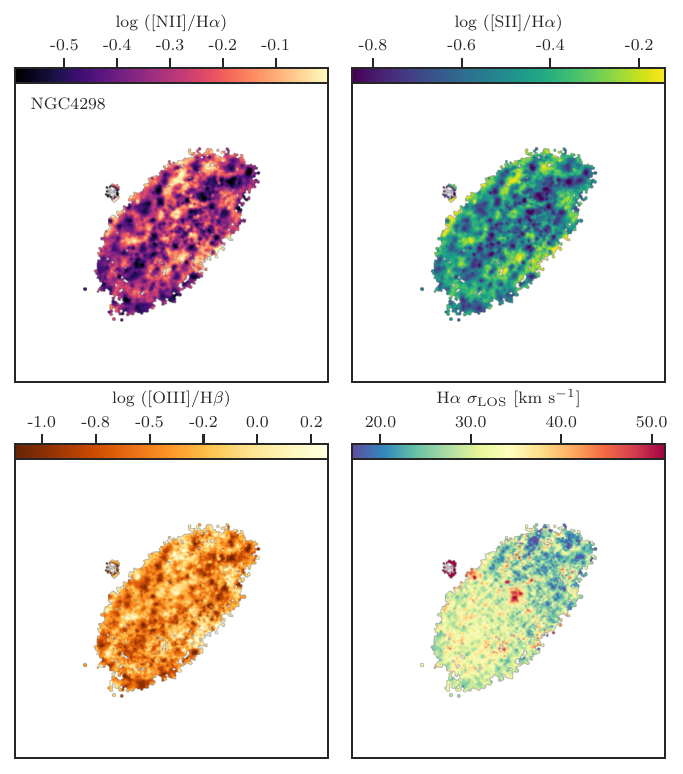}
    \caption{Same as Figure \ref{fig:NGC4501_emline_ratios} but for NGC~4298.}
\end{figure*}

\begin{figure*}
    \centering
    \includegraphics{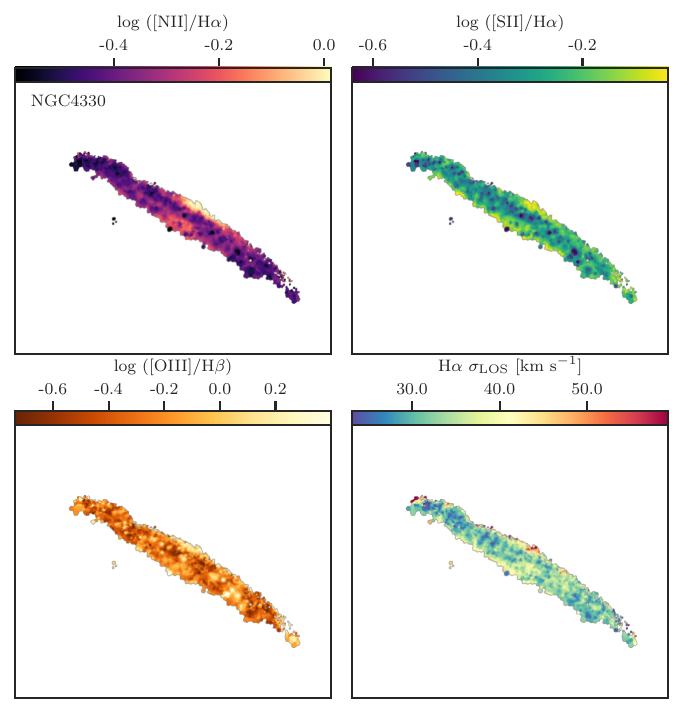}
    \caption{Same as Figure \ref{fig:NGC4501_emline_ratios} but for NGC~4330.}
\end{figure*}

\begin{figure*}
    \centering
    \includegraphics{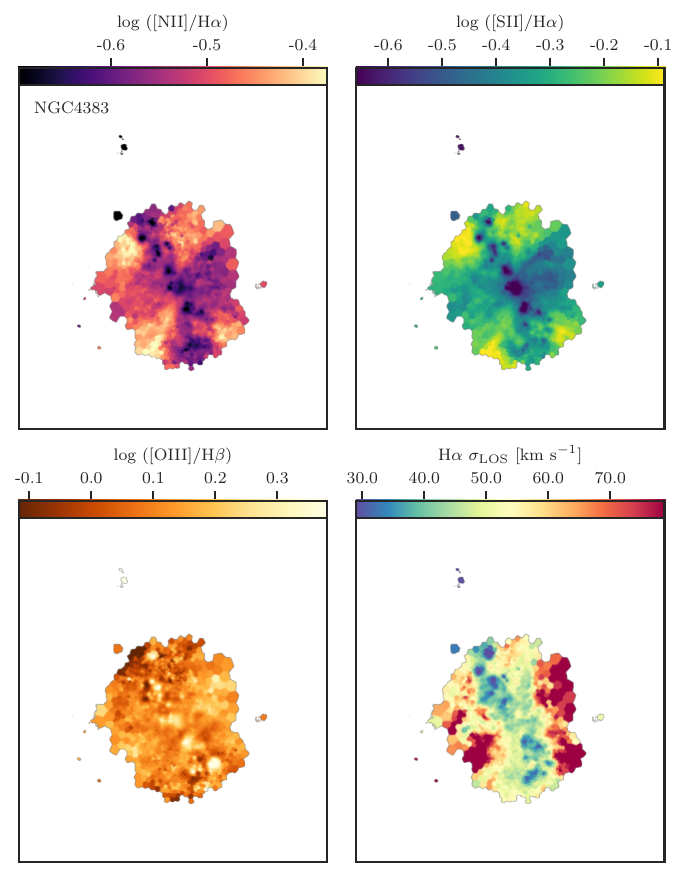}
    \caption{Same as Figure \ref{fig:NGC4501_emline_ratios} but for NGC~4383.}
\end{figure*}

\begin{figure*}
    \centering
    \includegraphics{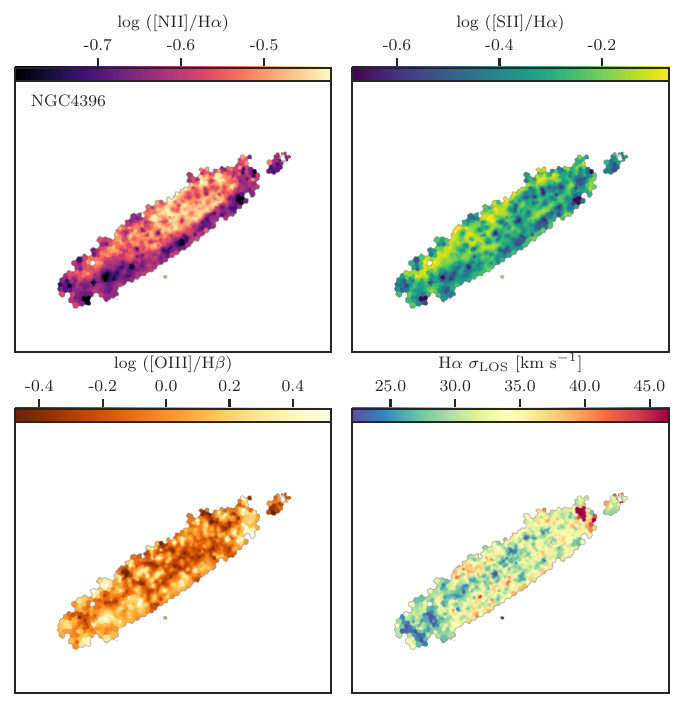}
    \caption{Same as Figure \ref{fig:NGC4501_emline_ratios} but for NGC~4396.}
\end{figure*}

\begin{figure*}
    \centering
    \includegraphics{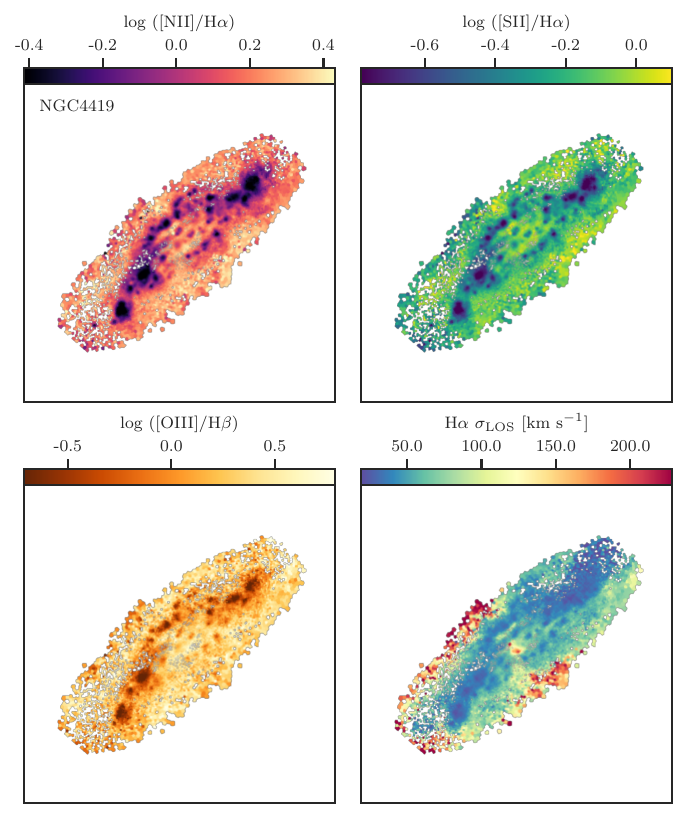}
    \caption{Same as Figure \ref{fig:NGC4501_emline_ratios} but for NGC~4419.}
\end{figure*}

\begin{figure*}
    \centering
    \includegraphics{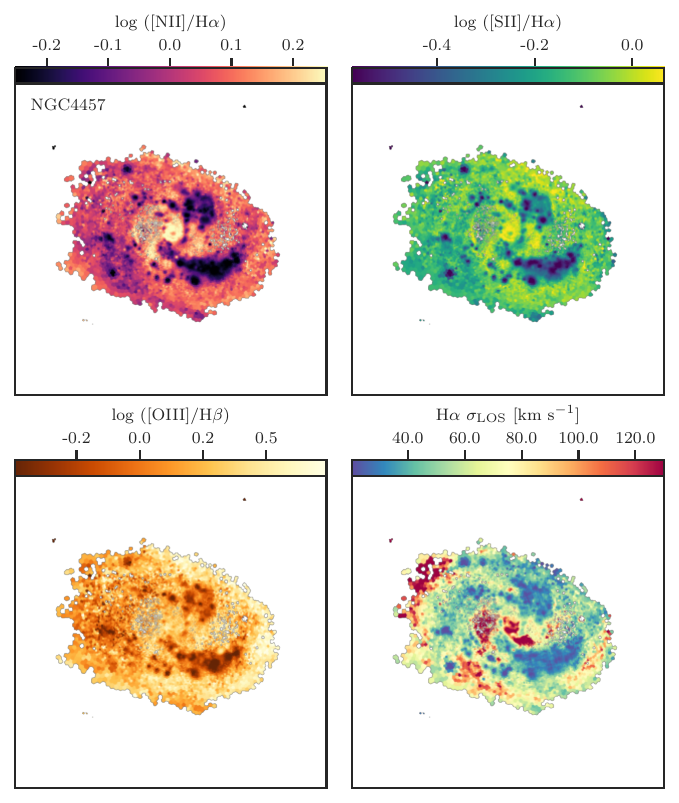}
    \caption{Same as Figure \ref{fig:NGC4501_emline_ratios} but for NGC~4457.}
\end{figure*}

\begin{figure*}
    \centering
    \includegraphics{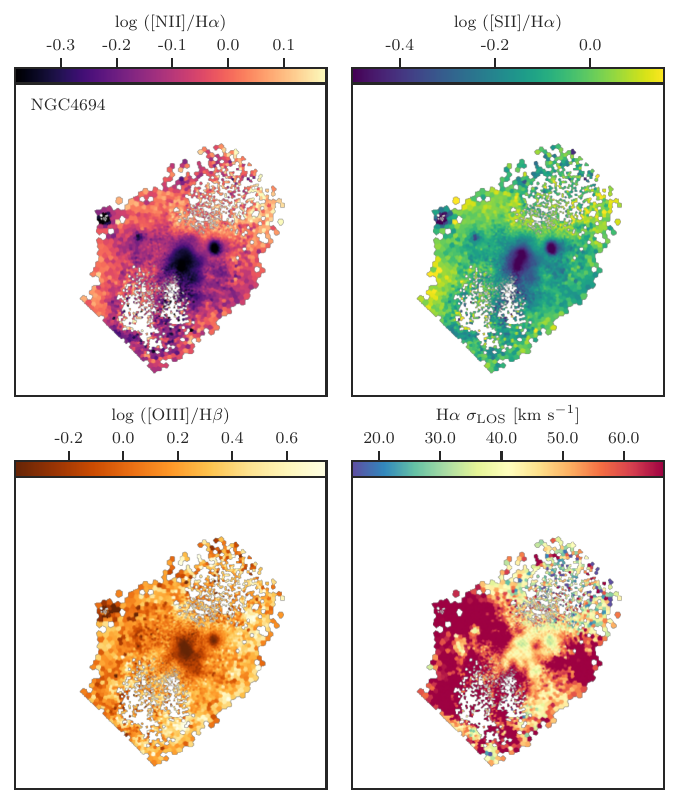}
    \caption{Same as Figure \ref{fig:NGC4501_emline_ratios} but for NGC~4694.}
\end{figure*}

\begin{figure*}
    \centering
    \includegraphics[height=0.9\textheight]{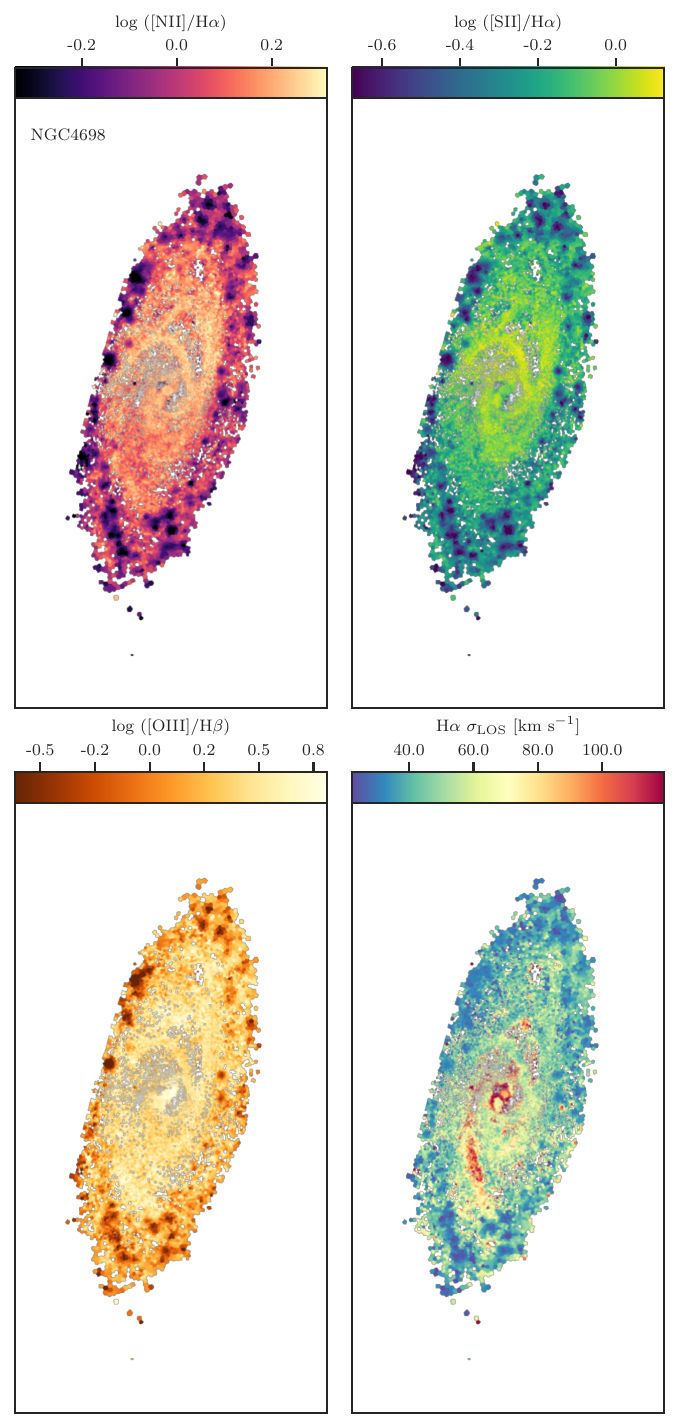}
    \caption{Same as Figure \ref{fig:NGC4501_emline_ratios} but for NGC~4698.}
\end{figure*}

\clearpage % This forces all pending figures to be printed NOW
\bibliography{refs}{}
\bibliographystyle{aasjournal}


\end{document}